\documentclass[a4paper,11pt]{article}

\usepackage{jheppub}
\usepackage{amsmath}
\usepackage{amssymb}
\usepackage{booktabs}
\usepackage{graphicx}
\usepackage{subcaption}

\graphicspath{{figures/}}

\begin{document}

\title{%
  Efficient negative-weight elimination in large high-multiplicity
  Monte Carlo event samples
}

\author[a]{Jeppe~R.~Andersen,}
\emailAdd{jeppe.andersen@durham.ac.uk}
\affiliation[a]{Institute for Particle Physics Phenomenology, Department of Physics, University of Durham, Durham, DH1 3LE, UK}

\author[b]{Andreas Maier}
\emailAdd{andreas.martin.maier@desy.de}
\affiliation[b]{Deutsches Elektronen-Synchrotron DESY, Platanenallee 6,
  15738 Zeuthen, Germany}

\author[a]{and Daniel Ma\^itre}
\emailAdd{daniel.maitre@durham.ac.uk}

\abstract{%
  We demonstrate that cell resampling can eliminate the bulk of
  negative event weights in large event samples of high multiplicity
  processes without discernible loss of accuracy in the predicted
  observables. The application of cell resampling to much larger data
  sets and higher multiplicity processes such as vector boson
  production with up to five jets has been made possible by improvements
  in the method paired with drastic enhancement of the computational
  efficiency of the implementation.
}

\begin{flushright}
IPPP/23/15, DCPT/23/30, DESY-23-037
\end{flushright}
\maketitle

\section{Introduction}
\label{sec:intro}

One of the greatest challenges in theoretical high-energy physics is
to meet the demand for increasingly precise predictions. Even leaving
aside conceptual issues, reducing theoretical uncertainties typically
requires ever more complex calculations, which incur steeply rising
computing costs. Already now Monte Carlo event generation for the LHC
constitutes a notable fraction of the experimental computing budgets.
Even with this large computing power investment event sample sizes
have to be limited to a size where the resulting uncertainty can be
non-negligible~\cite{HSFPhysicsEventGeneratorWG:2020gxw}. What is
more, event generation is only the first step in the full simulation
chain, and the already substantial computing costs of this step are
often dwarfed by the subsequent simulation of the detector
response. All these problems are expected to become even more severe
in the future, in particular with the advent of the HL-LHC. This
development is obviously at odds with general sustainability
goals. Improving the efficiency of event simulation is one of the
foremost tasks in particle physics phenomenology.

For a high-accuracy simulation a Monte Carlo generator needs to
combine real and virtual corrections, match resummations in different
limits and combine processes of different multiplicities while
avoiding double counting. Many different prescriptions exist for each
of these combination steps~\cite{Frixione:1995ms,Catani:1996vz,Catani:2002hc,Nagy:2003qn,Gehrmann-DeRidder:2005btv,Czakon:2010td,Somogyi:2005xz,Somogyi:2006da,Gaunt:2015pea,Cacciari:2015jma,Bonciani:2015sha,Magnea:2018hab,Frixione:2002ik,Frixione:2007vw,Hamilton:2012np,Hoche:2014uhw,Jadach:2015mza,Monni:2019whf,Prestel:2021vww,Catani:2001cc,Lonnblad:2001iq,Frederix:2012ps,Lonnblad:2012ix}. While they differ
substantially in the details, a common theme is the introduction of a
varying number of auxiliary events that \emph{subtract} from the
accumulated cross section instead of adding to it. If the number of
such negative-weight events becomes large enough, they can severly
impair the statistical convergence since large amounts of events are
required for a sufficient precision to allow for accurate
cancellation of the contributions with opposite signs. In fact, for a fractional
negative-weight contribution $r_-$ the number $N(r_-)$ of
required unweighted Monte Carlo events to reach a given statistics goal is (see e.g.~\cite{Frederix:2020trv})
\begin{equation}
  \label{eq:N_r_-}
  N(r_-) = \frac{N(0)}{(1 - 2r_-)^2}.
\end{equation}
Requiring a larger number of events not only increases the
computational cost in the generation stage, but especially also in the
subsequent detector simulation. A further problem is the increased disk
space usage, inducing both short- and long-term storage costs. It is
therefore highly desirable to keep the negative-weight contribution
small, $r_- \ll \frac{1}{2}$.

One avenue in this direction is to reduce the number of
negative-weight events during event generation, see
e.g.~\cite{Frederix:2020trv,Gao:2020vdv,Bothmann:2020ywa,Gao:2020zvv,Danziger:2021xvr}. A
second approach is to eliminate negative weights in the generated
sample, before detector
simulation~\cite{Andersen:2020sjs,Nachman:2020fff,Verheyen:2020bjw,Andersen:2021mvw}. In
the following, we focus on the \emph{cell resampling} approach
proposed in \cite{Andersen:2021mvw}, which in turn was inspired by
\emph{positive resampling}~\cite{Andersen:2020sjs}. Cell resampling is
independent of both the scattering process under consideration and any
later stages of the event simulation chain. It only redistributes event
weights, exactly preserving event kinematics. The effective range over
which event weights are smeared decreases with an increasing event
density. This implies that the smearing uncertainty decreases
systematically with increasing statistics without the need to change
the method. As demonstrated for the example
of the production of a W boson with two jets at next-to-leading-order
(NLO), a large fraction of negative weights can be eliminated without
discernible distortion of any predicted observable. A limitation of the original
formulation is the computational cost, which rises steeply with both the
event sample size and the number of final-state particles.

In section~\ref{sec:improvements}, we briefly review the method and
describe a number of algorithmic improvements which allow us to
overcome the original limitations through a speed-up by several orders
of magnitude. In section~\ref{sec:analyses}, we then apply cell
resampling to high-multiplicity samples with up to several billions of
events for the production of a W or Z boson in association with up to
five jets at NLO.  We conclude in section~\ref{sec:conclusions}.


\section{Algorithmic Improvements}
\label{sec:improvements}

In the following, we briefly review cell resampling and describe the
main improvements that allow us to apply the method also to large
high-multiplicity Monte Carlo samples. For a detailed motivation and
description of the original algorithm see~\cite{Andersen:2021mvw}.

\subsection{Cell Resampling}
\label{sec:cellres}

At its heart, cell resampling consists of repeatedly selecting a
subset of events --- referred to as cells --- and redistributing event
weights within the selected set. The steps are as follows.
\begin{enumerate}
\item Select an event with negative weight as the first event (the
``seed'') of a new cell $\mathcal{C}$.
\item Out of all events outside the cell, add the one with the
smallest distance from the seed to the cell. Repeat for as long as the
accumulated weight of all events within the cell is negative.
\item Redistribute the weights of the events inside the cell such that
the accumulated weight is preserved and none of the final weights is
negative.\footnote{A specific method of redistributing weights in this way is discussed in~\cite{Andersen:2021mvw}.}
\item Start anew with the next cell, i.e.\ with step 1.
\end{enumerate}
Note that an event can be part of several cells, but will only be
chosen as a cell seed at most once. In practice, we usually want to
limit the maximum cell size and abort step 2 once the distance between
the cell seed and its nearest neighbour outside the cell becomes too
large. We denote the maximum cell radius, i.e. the maximum allowed
distance between the cell seed and any other event within the cell, by
$d_\text{max}$. It is a parameter of the cell resampling algorithm.
If we limit the cell size in this way, we can only achieve a partial cancellation
between negative and positive event weights in the following step
3. For practical applications this is often sufficient, since the
small remaining contribution from negative weights has a much reduced
impact on the statistical convergence, cf.~equation~(\ref{eq:N_r_-}).

The computational cost of cell resampling tends to be completely
dominated by the nearest-neighbour search in step 2. In a naive
approach, one has to calculate the distances between the cell seed and
each other event in the sample. Since the number of cells is
proportional to the sample size $N$, the total computational
complexity is $\mathcal{O}(N^2)$. This renders the naive approach
unfeasible for samples with more than a few million events. For this reason,
an alternative approximate nearest-neighbour search based on
locality-sensitive hashing (LSH)~\cite{Indyk1998,Leskovec:2020} was
considered in~\cite{Andersen:2021mvw}. While this lead to an improved
scaling behaviour, the quality of the approximate search was also
found to deteriorate with an increasing sample size. An improved
version of this algorithm, discussed in appendix~\ref{sec:LSH_search}, still appears
to suffer from the same problem. In section~\ref{sec:nearest-neighbour_search}, we introduce an
exact search algorithm that is orders of magnitude faster than the
naive search.

The problem of costly distance calculations is further exacerbated by
the fact that a direct implementation of the originally proposed
distance function suffers from poor scaling for high
multiplicities. To compute the distance between two events $e$ and
$e'$, we first cluster the outgoing particles into infrared-safe
physics objects, e.g.~jets. We collect objects of the same type $t$
into sets $s_t$ for $e$ and $s_t'$ for $s$. The distance between the
two events is then
\begin{equation}
  \label{eq:d_event}
  d(e,e') = \sum_t d(s_t, s_t'),
\end{equation}
where $d(s_t, s_t')$ is the distance between the two sets $s_t, s_t'$. It is given by
\begin{equation}
  \label{eq:d_set}
  d(s_t,s_t') = \min_{\sigma \in S_P} \sum_{i=1}^P d(p_i, q_{\sigma(i)})\,,
\end{equation}
where $p_1,\dots, p_P$ are the momenta of the objects in $s_t$ and
$q_1,\dots, q_P$ the momenta\footnote{If the number of objects in $s_t$
and $s_t'$ is different, we add auxiliary objects with vanishing
momenta as described in~\cite{Andersen:2021mvw}.} in $s_t'$. A naive
minimisation procedure considers all permutations $\sigma$ in the symmetric
group $S_P$, i.e.~$P!$ possibilities. For large multiplicities $P$ a direct
calculation quickly becomes prohibitively
expensive. In~\cite{Andersen:2021mvw}, it was therefore suggested to
use an approximate scheme in this case. In
section~\ref{sec:set-to-set-distance} we discuss how the set-to-set
distance can be calculated both exactly and efficiently.

\subsection{Nearest-Neighbour Search}
\label{sec:nearest-neighbour_search}

Our improved nearest-neighbour search is based on vantage-point
trees~\cite{UHLMANN1991175,10.5555/313559.313789}. To construct a
vantage-point tree, we choose a single event as the first vantage
point. We then compute the distance to the vantage point for each
event. The closer half of the events lie within a hypersphere with
radius given by the median distance to the vantage point. We call the
populated part of this hypersphere the \emph{inside region} and its
complement the \emph{outside region}. We then recursively construct
vantage-point trees inside each of the two regions. The construction
terminates in regions that only contain a single point.

To find the nearest neighbour for any event $e$, we start at the root
of the tree, namely the first chosen vantage point. We calculate the
distance $D$ between this vantage point and $e$. If $D$ is less than
the radius $R$ of the hypersphere defining the inside region, we first continue the search in
the inside subtree, otherwise we choose the outside subtree
first. Let us first consider the case that the inside region is
the preferred one. It will contain a nearest-neighbour \emph{candidate} with
a distance $d$ to the initial event $e$. By the triangle inequality we
deduce that the \emph{actual} nearest neighbour can have a distance of at
most $D+d$ to the current vantage point. Therefore, if $D+d < R$, the actual
nearest neighbour cannot be in the outside region. Conversely, if
we started our search in the outside region and found a
nearest-neighbour candidate with $D-d > R$, then the actual nearest
neighbour cannot lie in the inside region. In summary, if $d < |R
- D|$ only the preferred region has to be considered.

Vantage-point tree search is indeed very well suited for cell
resampling. The construction is completely agnostic to the chosen
distance function. In particular, unlike the LSH-based methods
considered in~\cite{Andersen:2021mvw} and appendix~\ref{sec:LSH_search}, it does not
require a Euclidean metric. For an event sample of size $N$, the tree
construction requires $\mathcal{O}(N \log N)$ steps and can be easily
parallelised. In the ideal case where only the preferred regions are
probed, each nearest-neighbour search requires $\log_2 N$ comparisons,
which again results in an overall asymptotic complexity of
$\mathcal{O}(N \log N)$. While this means that for sufficiently large
event samples cell resampling will eventually require more computing
time than the $\mathcal{O}(N)$ event generation, we find that this is
not the case for samples with up to several billion events. Timings
for a practical application are given in section~\ref{sec:timings}.

We further optimise the nearest-neighbour search in several
aspects. Most importantly, if we limit the maximum cell size to $d_{\text{max}}$,
we can dramatically increase the probability that only the preferred
regions have to be considered. In fact, if $|R - D| > d_{\text{max}}$ then any
suitable nearest neighbours have to lie inside the preferred
region. We can further enhance the probability through a judicious choice
of the vantage points. Since input events near the boundary between
inside and outside regions require checking both regions for
nearest neighbours, the general goal is to minimise this surface. To
this end, we choose our first vantage point at the boundary of the
populated phase space. We select a random event, calculate
the distance to all other events, and choose the event with the
largest distance as the vantage point. Then, when constructing the
subtrees for the inside and outside regions, we choose as
vantage points those events that have the largest distance to the
parent vantage point.

When constructing a cell, we have to find nearest neighbours until
either the accumulated weight becomes non-negative or the distance
exceeds the maximum cell radius. This corresponds to a so-called $k$
nearest neighbour search, where in this case the required number $k$
of nearest neighbours is a priori unknown. To speed up successive
searches, we cache the results of distance calculations, i.e.~all
values of $D$ for a given input event.

Finally, we note that the vantage-point tree can also be employed for
approximate nearest-neighbour search if one only searches the preferred
region in each step. We exploit this property by first partitioning
the input events into the inside and outside regions of a
shallow vantage point tree, aborting the construction already after
the first few steps. We then apply cell resampling to each partition
independently. This approach allows efficient parallelisation, while
yielding much better results than the independent cell resampling of
randomly chosen partial samples. The price to pay is that the quality
of the nearest-neighbour search and therefore also of the overall
resampling deteriorates to some degree. In practice this effect
appears to be minor, see also section~\ref{sec:weight_elim}.

\subsection{Set-to-Set Distance at High Multiplicities}
\label{sec:set-to-set-distance}

The distance between two events as defined in~\cite{Andersen:2021mvw}
is the sum of distances between sets of infrared-safe physics objects,
see equation~\eqref{eq:d_event}. To define the distance between two
such sets $s_t, s_t'$, we aim to find the optimal pairing between the
momenta $p_1,\dots, p_P$ of the objects in $s_t$ and the momenta $q_1,\dots, q_P$ of the objects in $s_t'$. The naive approach of
considering all possible pairings, cf.~equation~\eqref{eq:d_set},
scales very poorly with the number of objects. However, the task of
finding an optimal pairing is an instance of the well-studied
\emph{assignment problem}.

Let us introduce the matrix $\mathbb{D}$ of distances with
\begin{equation}
  \label{eq:dist_matrix}
  \mathbb{D}_{ij} \equiv d(q_i, p_j).
\end{equation}
An efficient method for minimising $\sum_{i=1}^{P}
\mathbb{D}_{i\sigma(i)}$ was first found by
Jacobi~\cite{Jacobi1,Jacobi2}. It was later rediscovered independently
and popularised under the name ``Hungarian
method''~\cite{Kuhn1,Kuhn2,Munkres,10.1145/321694.321699,https://doi.org/10.1002/net.3230010206}. The
algorithm mutates the entries of $\mathbb{D}_{ij}$ in such a way that
the optimal pairing is preserved during each step. After each step,
one marks a minimum number of rows and columns such that each
vanishing entry is part of a marked row or column. The algorithm
terminates as soon as all rows (or columns) have to be marked. The mutating steps are as follows:
\begin{enumerate}
\item Replace each $\mathbb{D}_{ij}$ by $\mathbb{D}_{ij} - \min_{k} \mathbb{D}_{ik}$.
\item Replace each $\mathbb{D}_{ij}$ by $\mathbb{D}_{ij} - \min_{k} \mathbb{D}_{kj}$.
\item Find the smallest non-vanishing entry. Subtract it from all
unmarked rows and add it to all marked columns.
\end{enumerate}
Step 3 is repeated until the termination criterion is fulfilled.  In
our code, we use the implementation in the
\texttt{pathfinding}~\cite{pathfinding} package. Like the remainder of
our implementation of cell resampling it is written in the Rust
programming language.

Using the Hungarian algorithm instead of a brute-force search improves
the scaling behaviour for sets with $P$ momenta from $\mathcal{O}(P!)$
to $\mathcal{O}(P^3)$. In practice, we find it superior for $P >
3$. The \emph{FlowAssign} algorithm proposed by Ramshaw and Tarjan
\cite{Ramshaw2012581} would scale even better, with a time complexity
of $\mathcal{O}\big(P^{5 / 2}\log(D P)\big)$. The caveat is that the
scaling also depends logarithmically on the range $D$ of distances
encountered. Since the maximum multiplicity reached in current NLO
computations is limited, an auction \cite{Bertsekas1988} (or
equivalently push-relabel \cite{Goldberg1995,Alfaro2022}) algorithm
may still perform better in practice despite formally inferior scaling
behaviour. We leave a detailed comparison to future work.


\section{Negative Weight Elimination in Vector Boson plus Jets Production at NLO}
\label{sec:analyses}

We are now in a position to apply cell resampling to large
high-multiplicity event samples. We consider the production of a
vector boson in association with jets at NLO, using
\textsc{ROOT}~\cite{Brun:1997pa} \texttt{ntuple}~\cite{Bern:2013zja,Anger:2017nkq}
event files generated with \textsc{BlackHat}~\cite{Berger:2008sj} and
\textsc{Sherpa} 2.1~\cite{Gleisberg:2008ta}. Jets are defined
according to the anti-$k_t$ algorithm~\cite{Cacciari:2008gp} with $R =
0.4$ and a minimum transverse momentum of $25\,$GeV. More details on
the event generation are given in~\cite{Bern:2013gka,Anger:2017nkq}. The various
samples with their most salient properties are listed in
table~\ref{tab:samples}.

\begin{table}[htb]
  \centering
  \begin{tabular}{llll}
    \toprule
    Sample & Process & Centre-of-mass energy & \# events \\
    \midrule
    \texttt{Z1} & $pp \to (Z \to e^+e^-) + \text{jet}$ & 13\,TeV & $8.21\times 10^8$\\
    \texttt{Z2} & $pp \to (Z \to e^+e^-) + 2\text{ jets}$ & 13\,TeV & $5.30\times 10^8$\\
    \texttt{Z3} & $pp \to (Z \to e^+e^-) + 3\text{ jets}$ & 13\,TeV & $1.65\times 10^9$\\
    \texttt{W5} & $pp \to (W^- \to e^- \nu_e) + 5\text{ jets}$ & 7\,TeV & $1.17 \times 10^9$\\
    \bottomrule
  \end{tabular}
  \caption{NLO event samples used for cell resampling.}
  \label{tab:samples}
\end{table}

We apply cell resampling to each of the samples, defining
infrared-safe physics objects according to the above jet
definition. We use the distance function defined in
\cite{Andersen:2020sjs}, which follows from
equations~\eqref{eq:d_event}, \eqref{eq:d_set}, and the momentum
distance
\begin{equation}
  \label{eq:p_dist}
  d(p, q) = \sqrt{\lvert\vec{p} - \vec{q}\,\rvert^2 + \tau^2 (p_\perp - q_\perp)^2}.
\end{equation}
Here, we set $\tau = 0$ and limit the maximum cell radius to 10\,GeV
for samples \texttt{Z1}, \texttt{Z2}, and \texttt{W5}, and to 2\,GeV
for sample \texttt{Z3}. To examine the impact of these choices for the
maximum radius we additionally compare to a resampling run with a
maximum cell size of 100\,GeV for sample \texttt{W5}.

For better parallelisation and general performance we
pre-partition each input sample into several regions according to one
of the upper levels of a vantage-point tree, as explained in
section~\ref{sec:nearest-neighbour_search}. Here, we use the seventh
level, corresponding to 128 regions.

To interpret our results we use standard
\textsc{Rivet}~\cite{Bierlich:2019rhm} analyses. We verify that the
event count and total cross section of each sample is preserved using
the \texttt{MC\_XS} analysis. Furthermore, we employ this analysis to
assess the degree to which negative weights are eliminated. For the
sample \texttt{W5} we additionally use the \texttt{MC\_WINC} and
\texttt{MC\_WJETS} analysis, and their counterparts \texttt{MC\_ZINC}
and \texttt{MC\_ZJETS} for the remaining samples involving a Z
boson. We investigate the impact of additional cuts applied after cell
resampling using the ATLAS analysis
\texttt{ATLAS\_2017\_I1514251}~\cite{ATLAS:2017sag} for inclusive Z
boson production.

\subsection{Comparison of Predictions}
\label{sec:comp_pred}

\begin{figure}[htb]
  \centering
  \includegraphics[width=0.45\linewidth]{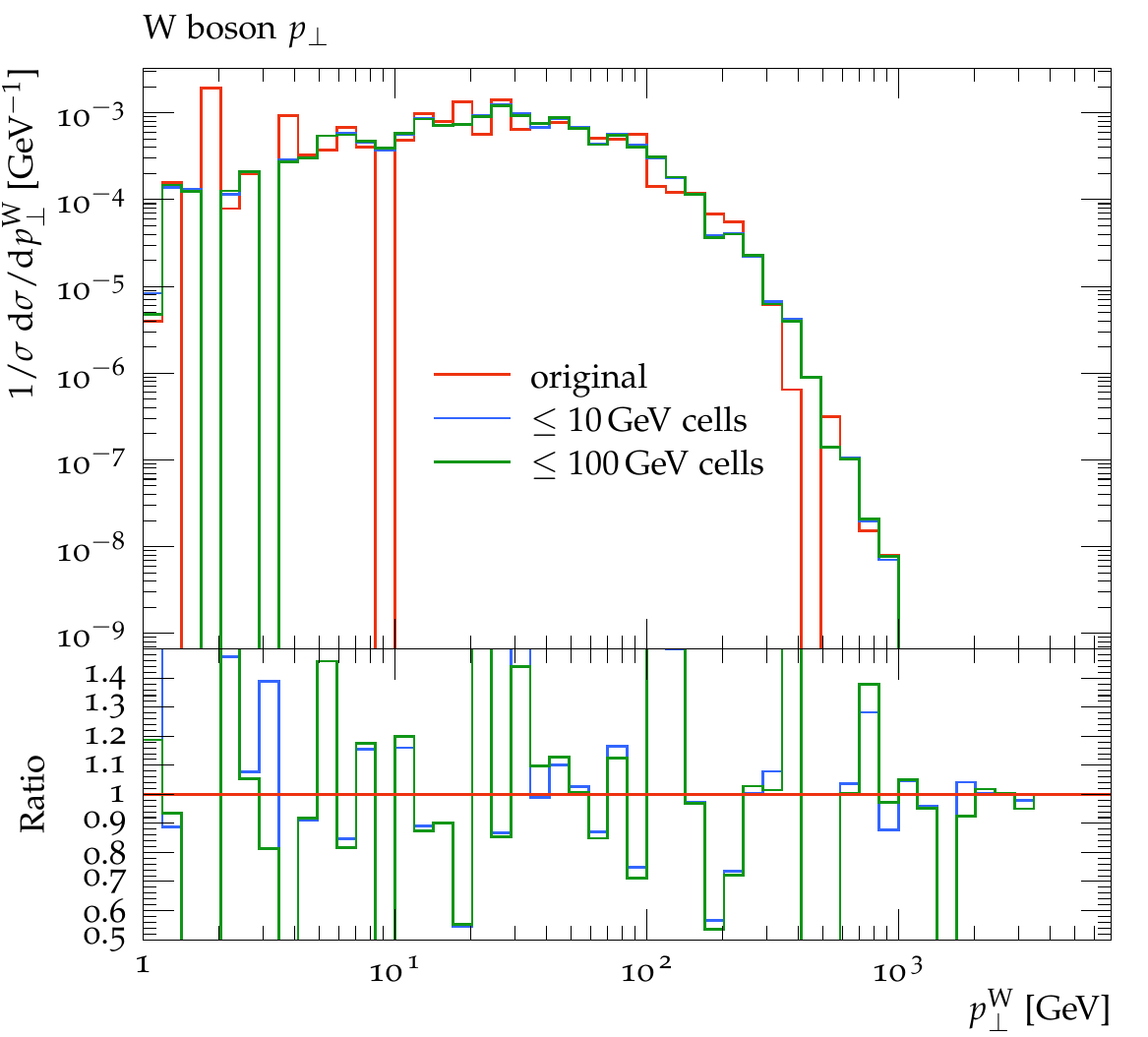}
  \includegraphics[width=0.45\linewidth]{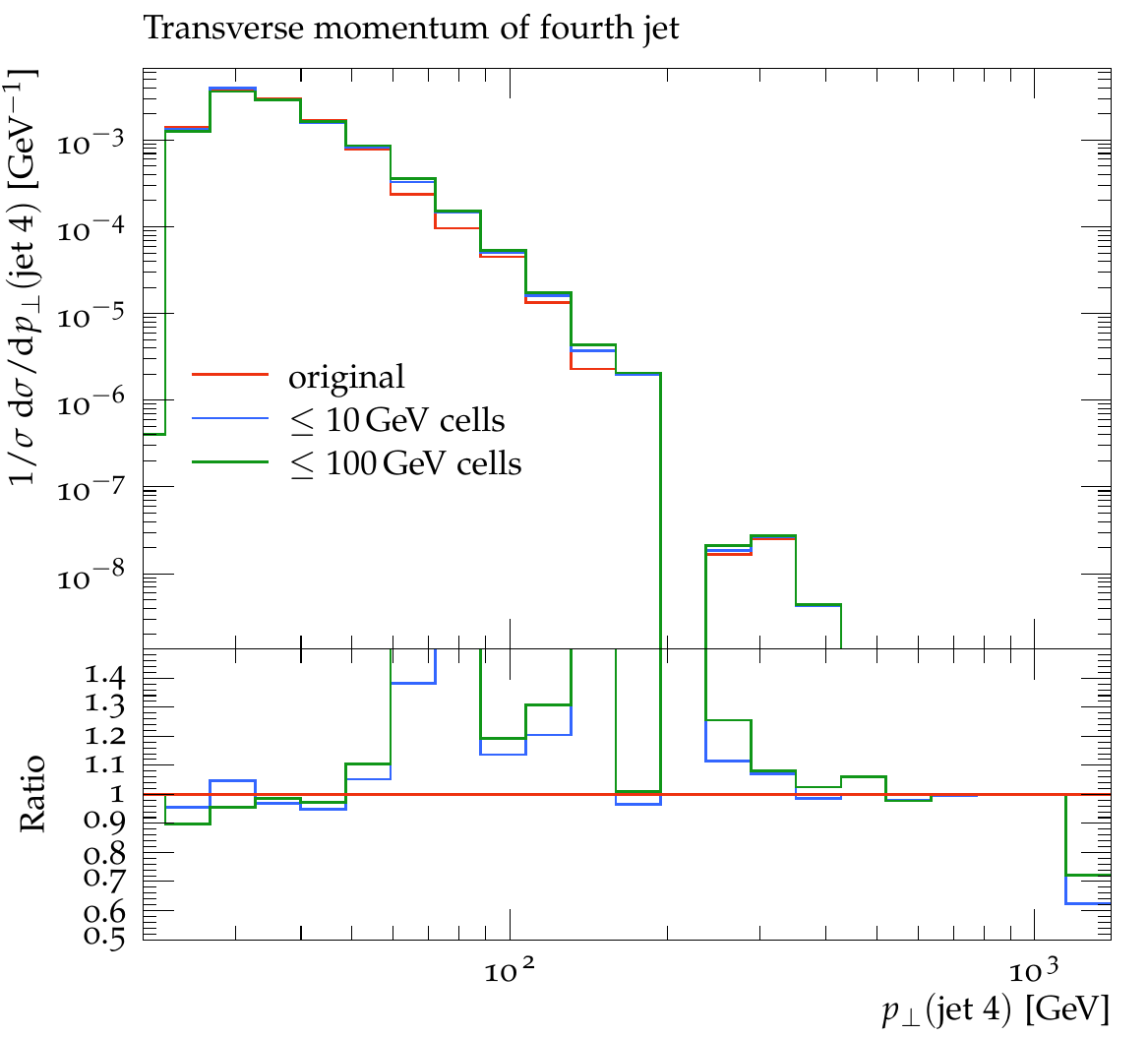}
  \includegraphics[width=0.45\linewidth]{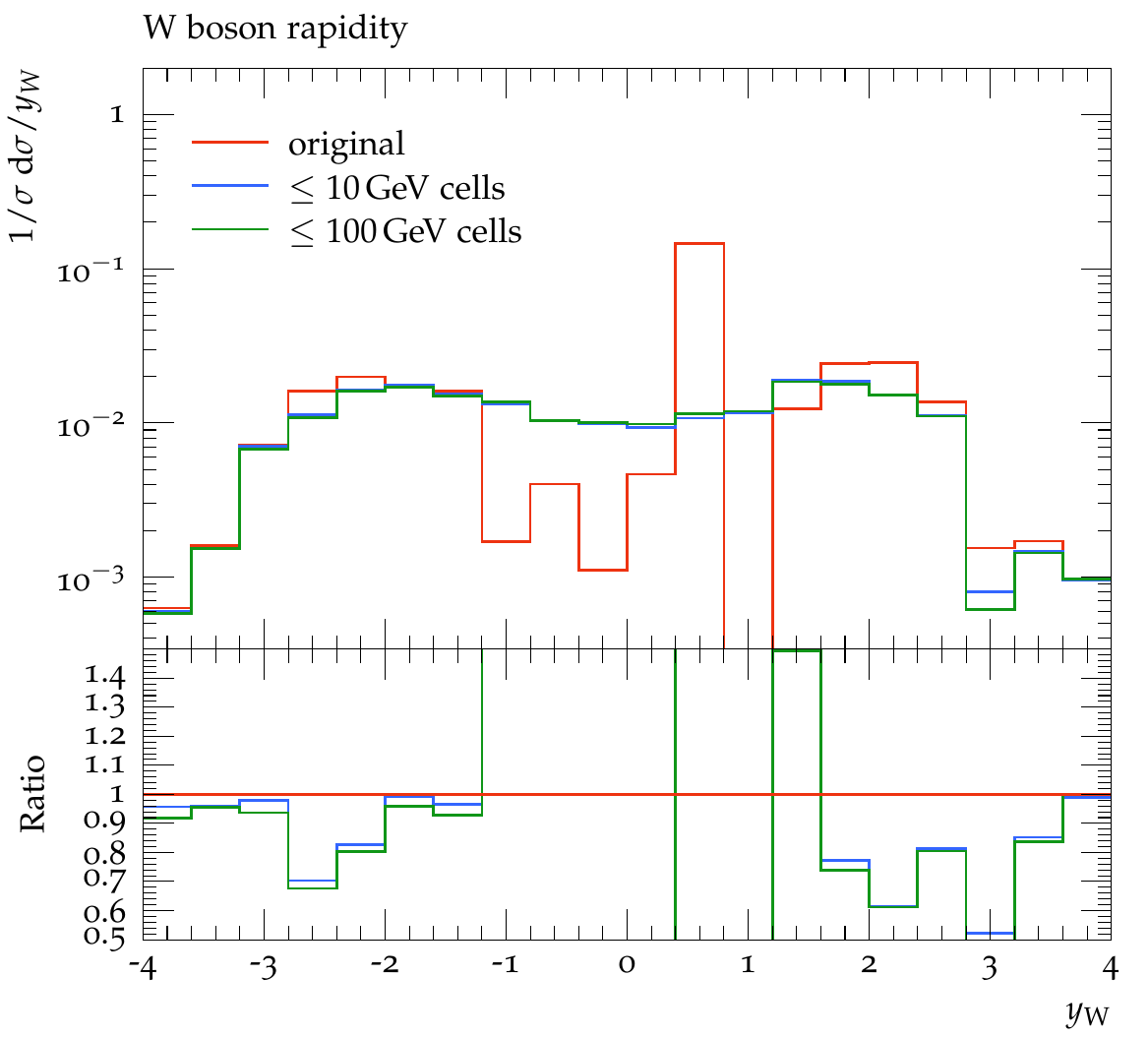}
  \includegraphics[width=0.45\linewidth]{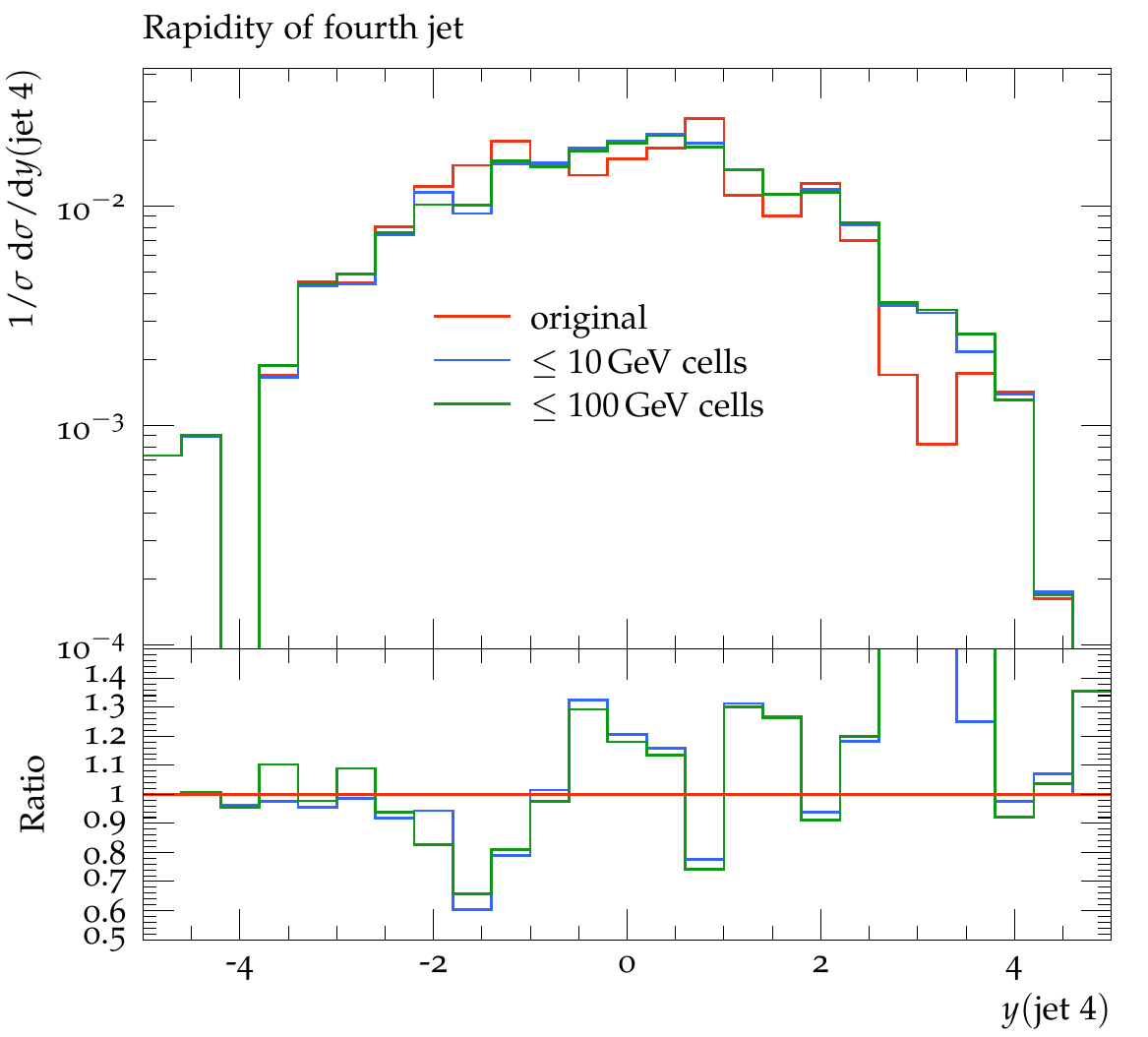}
  \includegraphics[width=0.45\linewidth]{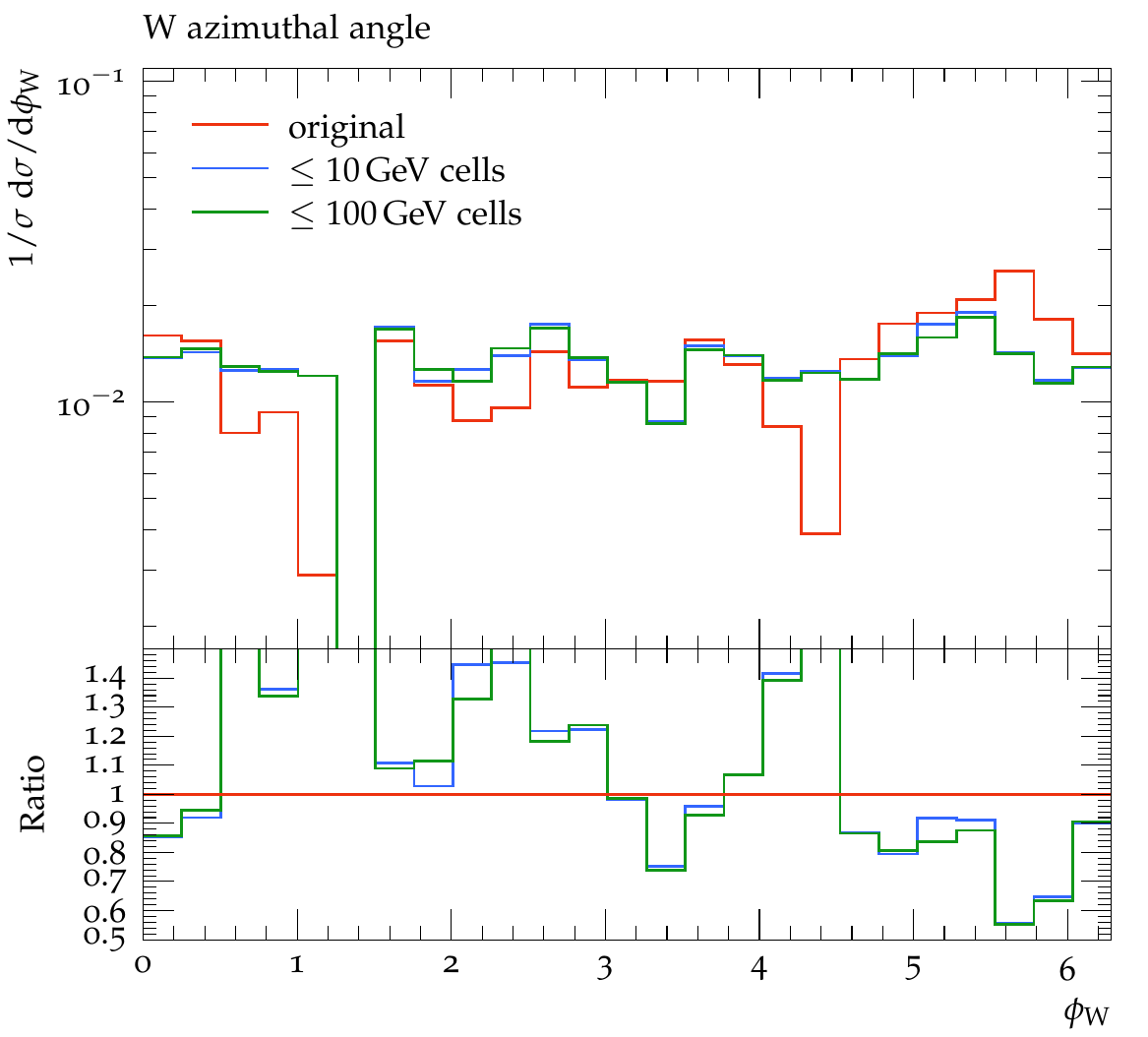}
  \includegraphics[width=0.45\linewidth]{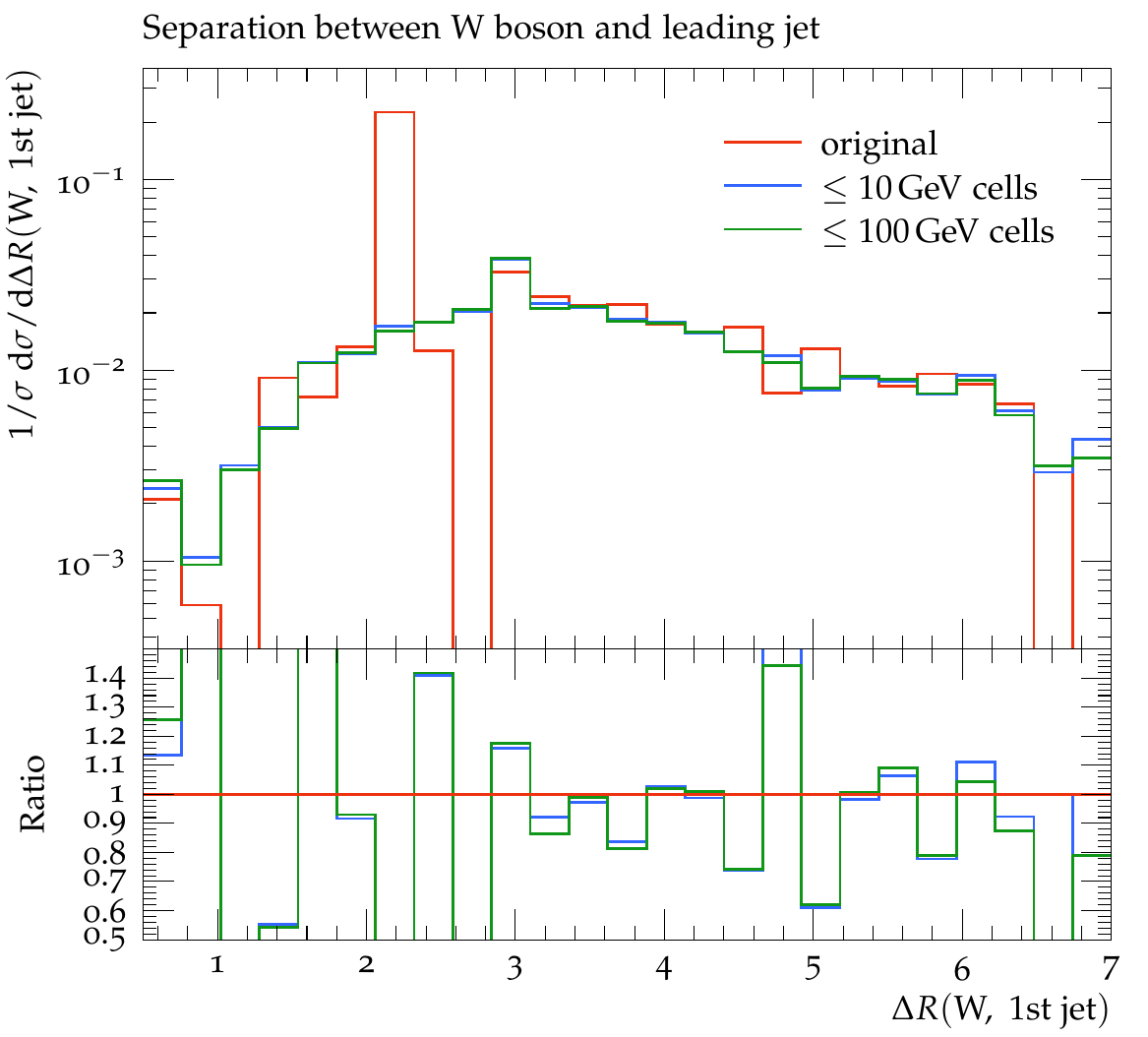}
  \caption{%
    Comparison of distributions before and after cell resampling for
    sample \texttt{W5} in table~\ref{tab:samples}. The blue lines indicate
    cell resampling with a maximum cell radius of 10\,GeV, the green lines
    result from a radius limit of 100\,GeV. Distributions are normalised
    according to the total cross section for sample \texttt{W5}.
  }
  \label{fig:comp_W5}
\end{figure}

\begin{figure}[htb]
  \centering
  \begin{subfigure}{0.45\linewidth}
    \label{fig:Z1_pt}
      \includegraphics[width=\linewidth]{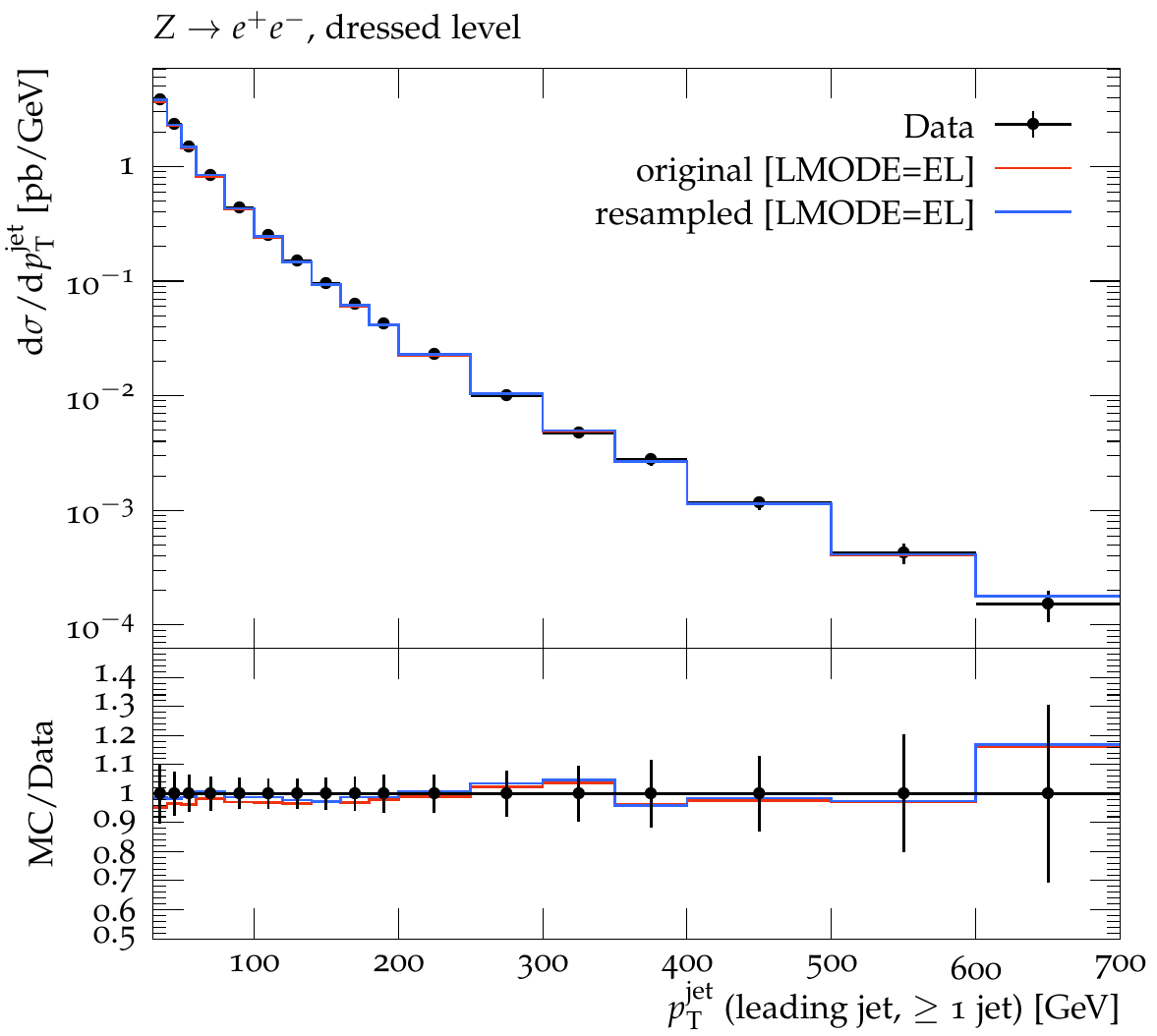}
      \caption{$p_\perp$ of hardest jet in sample \texttt{Z1}}
    \end{subfigure}
\begin{subfigure}{0.45\linewidth}
    \label{fig:Z2_pt}
    \includegraphics[width=\linewidth]{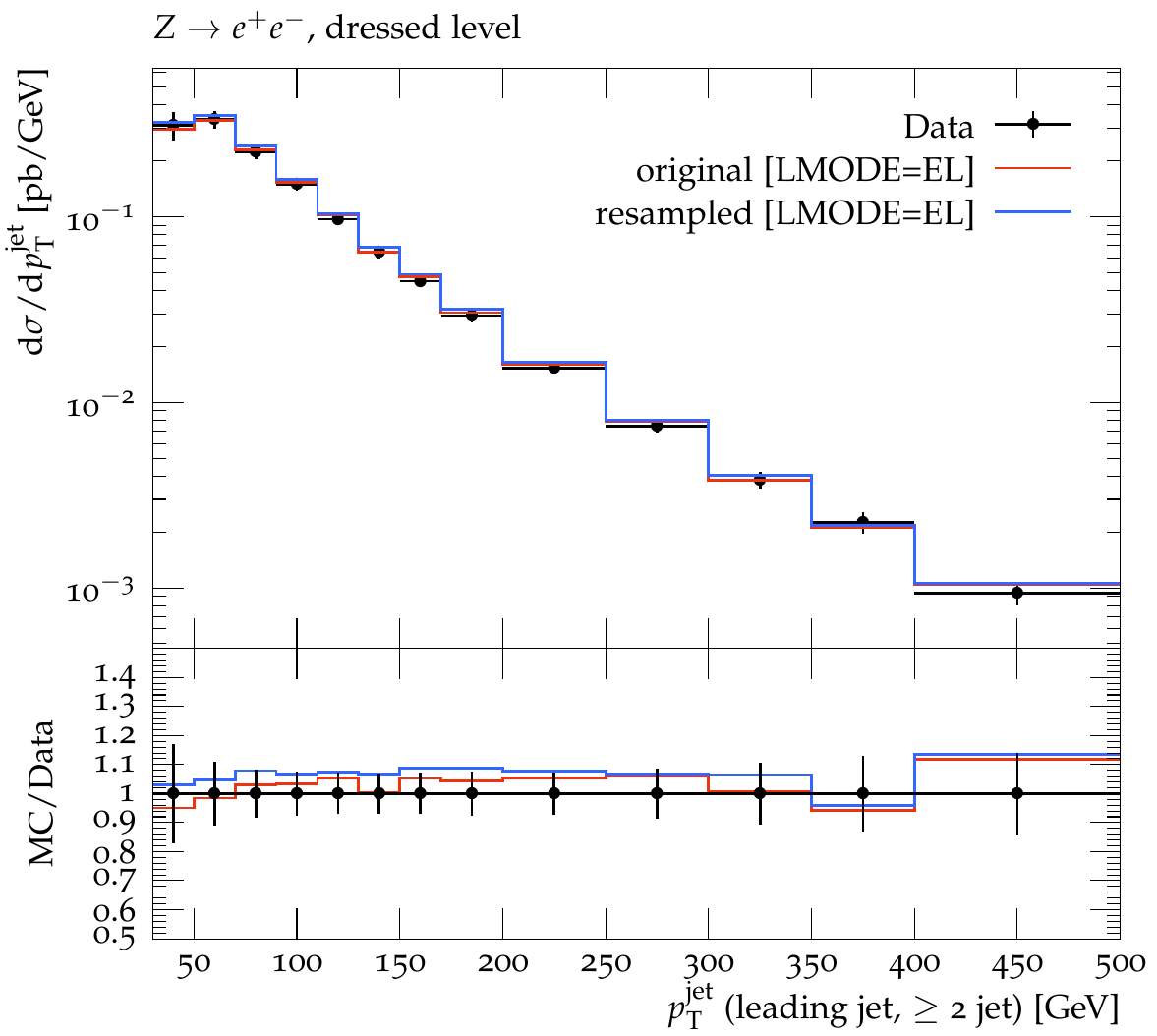}
  \caption{$p_\perp$ of hardest jet in sample \texttt{Z2}}
\end{subfigure}
\begin{subfigure}{0.45\linewidth}
      \label{fig:Z3_pt}
  \includegraphics[width=\linewidth]{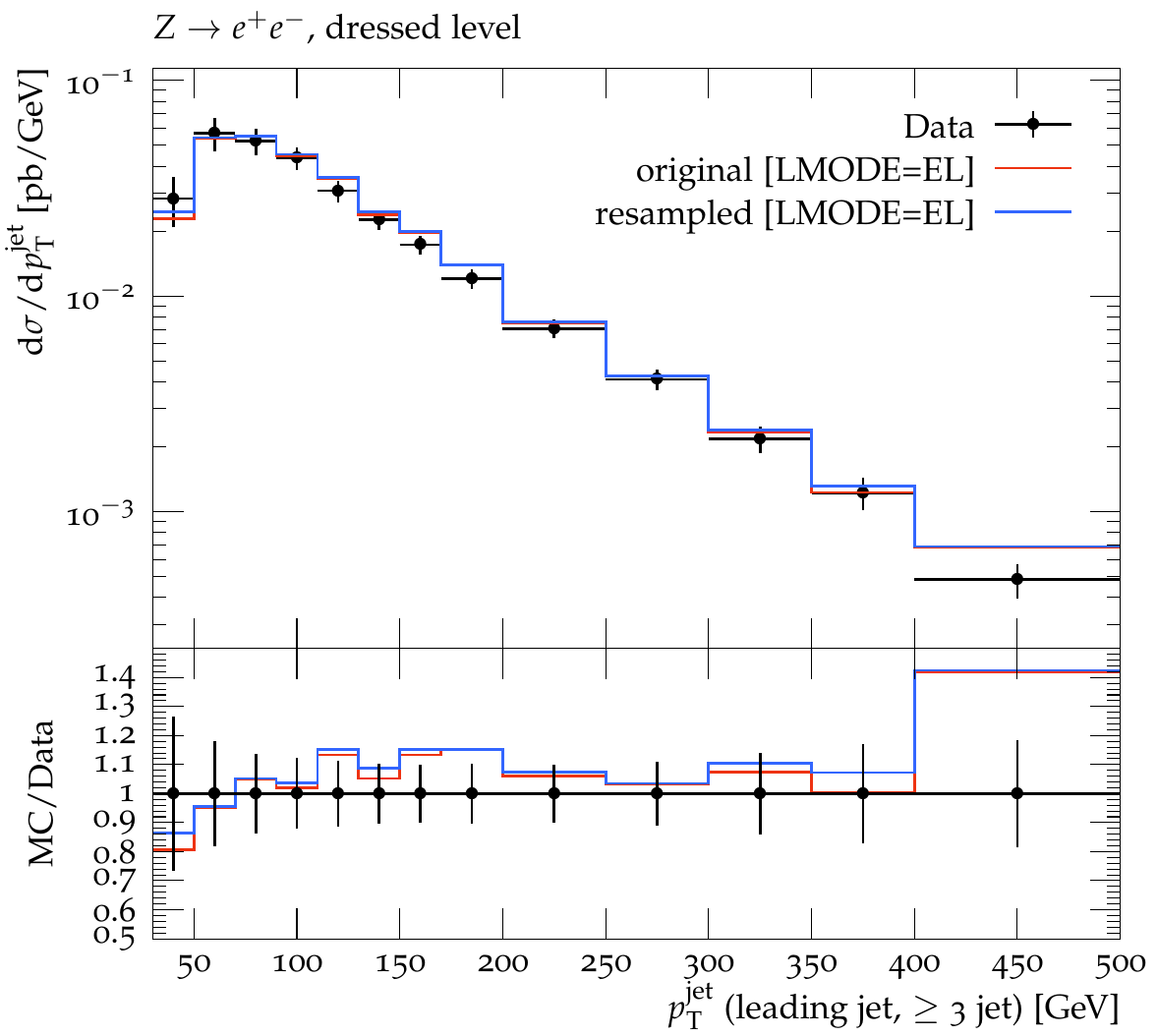}
  \caption{$p_\perp$ of hardest jet in sample \texttt{Z3}}
\end{subfigure}
\begin{subfigure}{0.45\linewidth}
      \label{fig:Z1_y}
  \includegraphics[width=\linewidth]{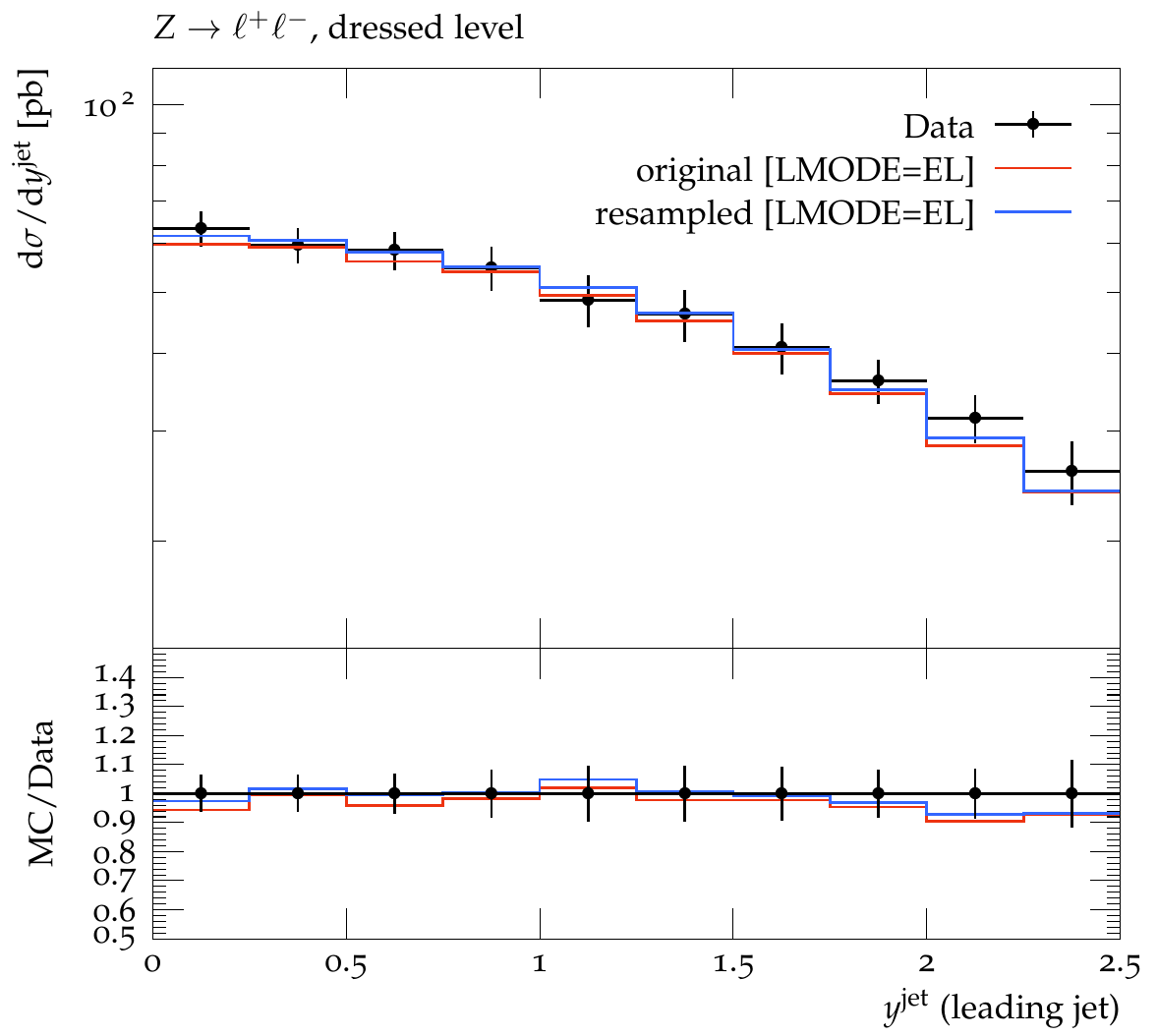}
  \caption{Rapidity of hardest jet in sample \texttt{Z1}}
\end{subfigure}
\begin{subfigure}{0.45\linewidth}
      \label{fig:Zn_y}
  \includegraphics[width=\linewidth]{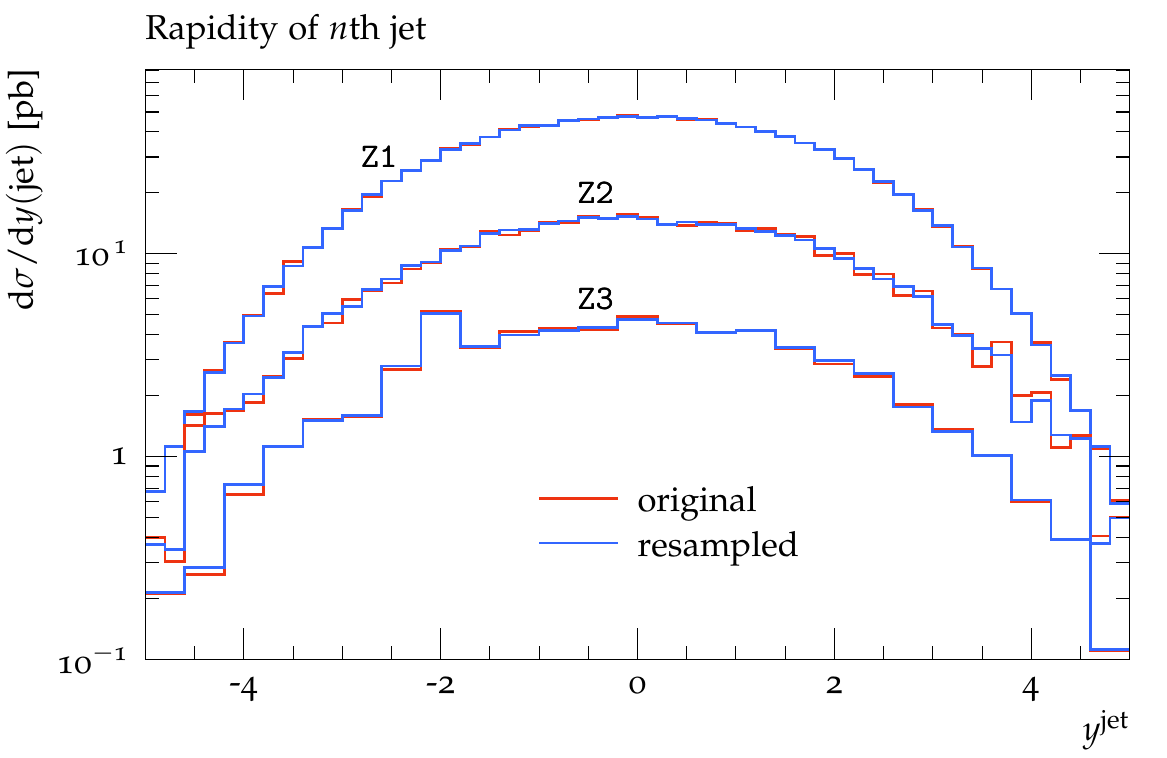}
  \caption{Rapidity of $n$th jet in sample \texttt{Z}$n$}
\end{subfigure}
\caption{%
  Comparison of distributions before and after cell resampling for
  samples \texttt{Z1}, \texttt{Z2}, and \texttt{Z3} in table~\ref{tab:samples}.
  Subfigures (a) to (d) show jet transverse momentum and rapidity
  distributions taken from the \texttt{ATLAS\_2017\_I1514251} Rivet
  analysis. Subfigure (e) is a jet rapidity distribution taken from
  \texttt{MC\_ZJETS}.
  }
  \label{fig:comp_Z1_to_Z3}
\end{figure}

We first assert that predictions remain equivalent by comparing a
number of distributions before and after cell
resampling. Figure~\ref{fig:comp_W5} shows a variety of distributions
for sample \texttt{W5}. In figure~\ref{fig:comp_Z1_to_Z3} we show selected
distributions for the \texttt{Z1}, \texttt{Z2}, and \texttt{Z3}
samples. In all cases, we find that the differences between original
and resampled predictions are comparable to or even smaller than the
statistical bin-to-bin fluctuations in the original. A more indirect
way to estimate the bias introduced by cell resampling is to consider
the characteristic cell radii and the spread of measured observables
within the cells. This is discussed in appendix \ref{sec:cell_sizes}.

\subsection{Improvement in Sample Quality}
\label{sec:weight_elim}

In order to assess the improvement achieved through cell resampling,
we first consider the reduction in the negative weight
contribution. To this end, we determine how much larger the original
and the resampled event samples have to be to reach the same
statistical power as an event sample without negative weights. In
other words, we compute $N(r_-)$ as defined in
equation~\eqref{eq:N_r_-}, where the fractional negative-weight
contribution $r_- = \sigma_-/(\sigma_+ + \sigma_-)$ is obtained from
the contribution $\sigma_+$ of positive-weight events to the total
cross section $\sigma$ and the absolute value of the negative-weight
cross section contribution $\sigma_- = \sigma_+ - \sigma$.

\begin{figure}[htb]
  \centering
  \includegraphics[width=\linewidth]{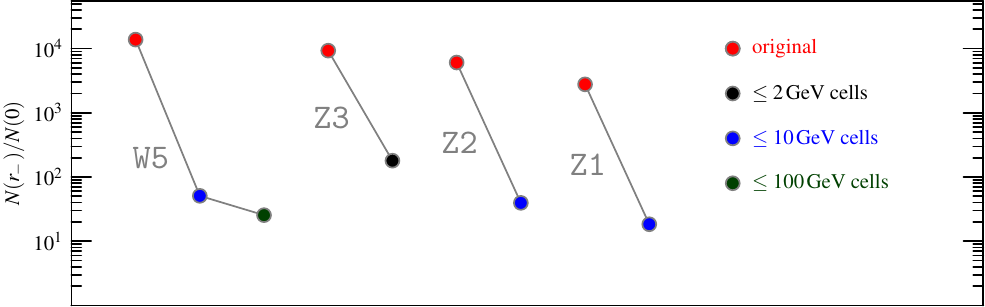}
  \caption{%
    Required number of events relative to an ideal event sample without
    negative weights before and after resampling. Event samples are
    labeled as listed in table~\ref{tab:samples}.
  }
  \label{fig:Neff}
\end{figure}

As demonstrated in figure~\ref{fig:Neff}, cell resampling leads to a
drastic improvement by roughly two to three orders of
magnitude. Increasing the maximum cell radius leads to an even
stronger reduction, at the cost of increased computing time and
potentially larger systematic errors introduced by the procedure. To
assess the impact of pre-partioning the event samples, we
alternatively resample \texttt{Z1} without prior partitioning. This
leads to a slight reduction of $N(r_-)/N(0)$ from 18.4 with
pre-partitioning to 17.1 without pre-partitioning.

\begin{figure}[htb]
  \centering
      \includegraphics[width=0.45\linewidth]{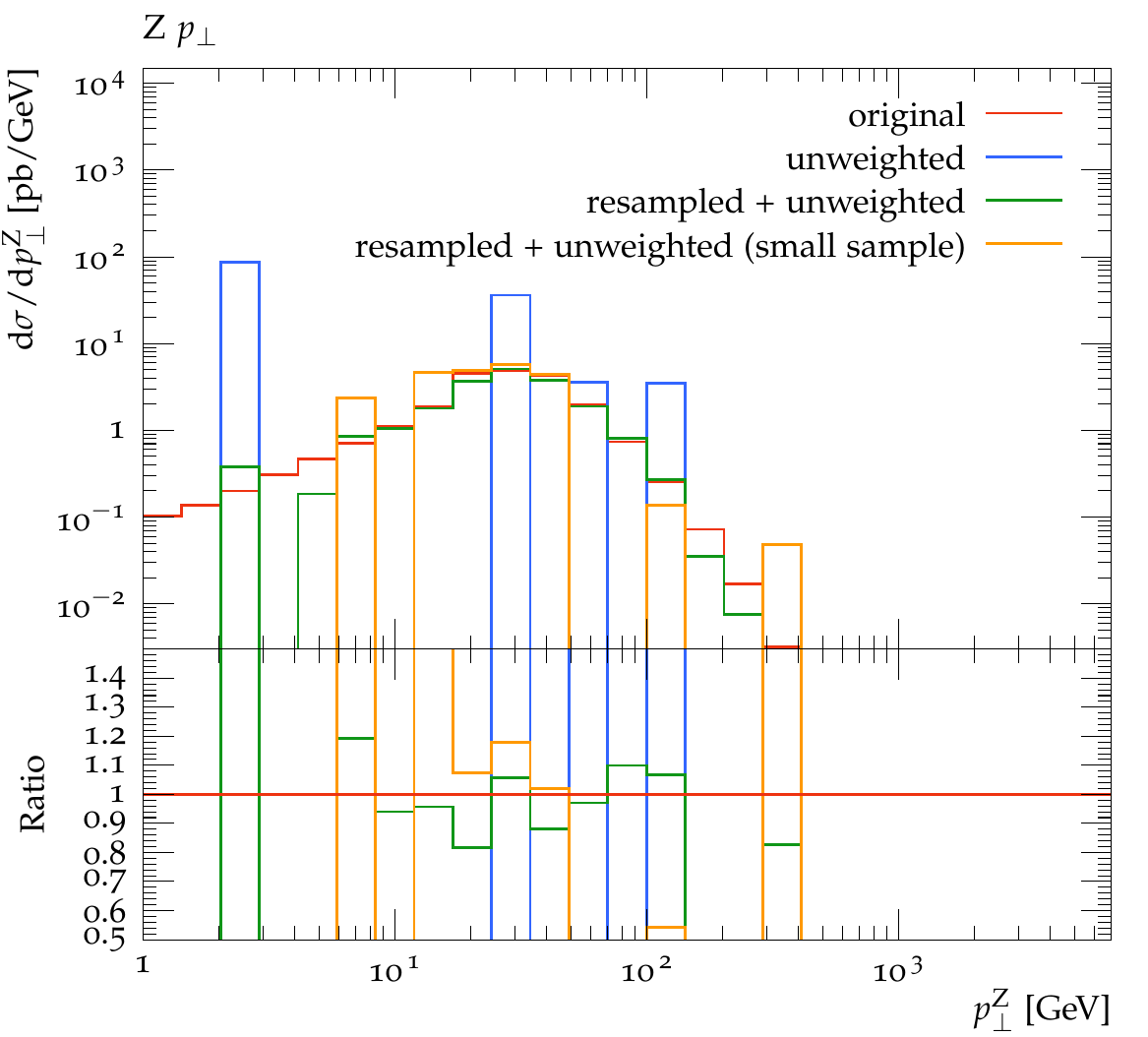}
      \includegraphics[width=0.45\linewidth]{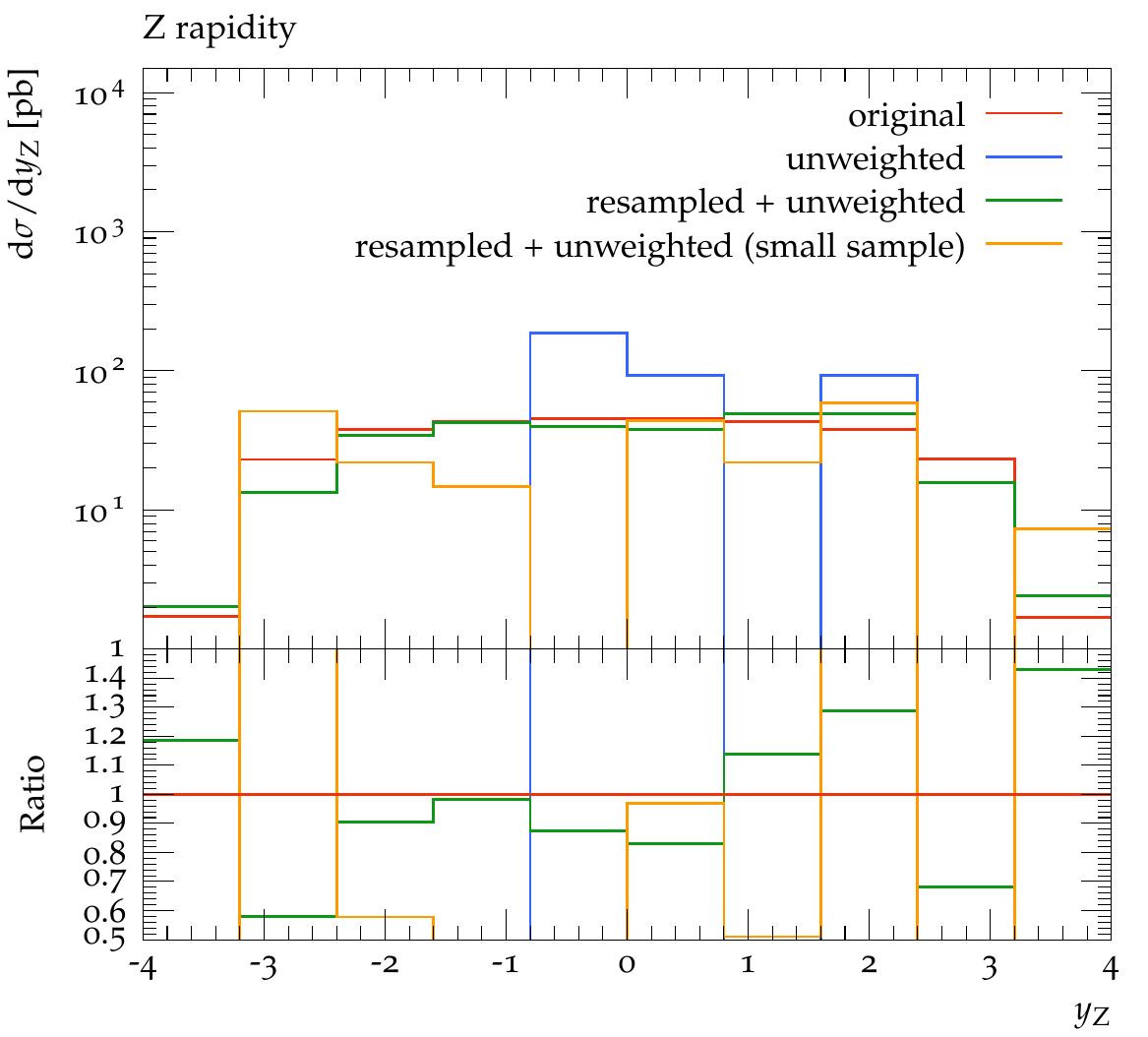}
      \includegraphics[width=0.45\linewidth]{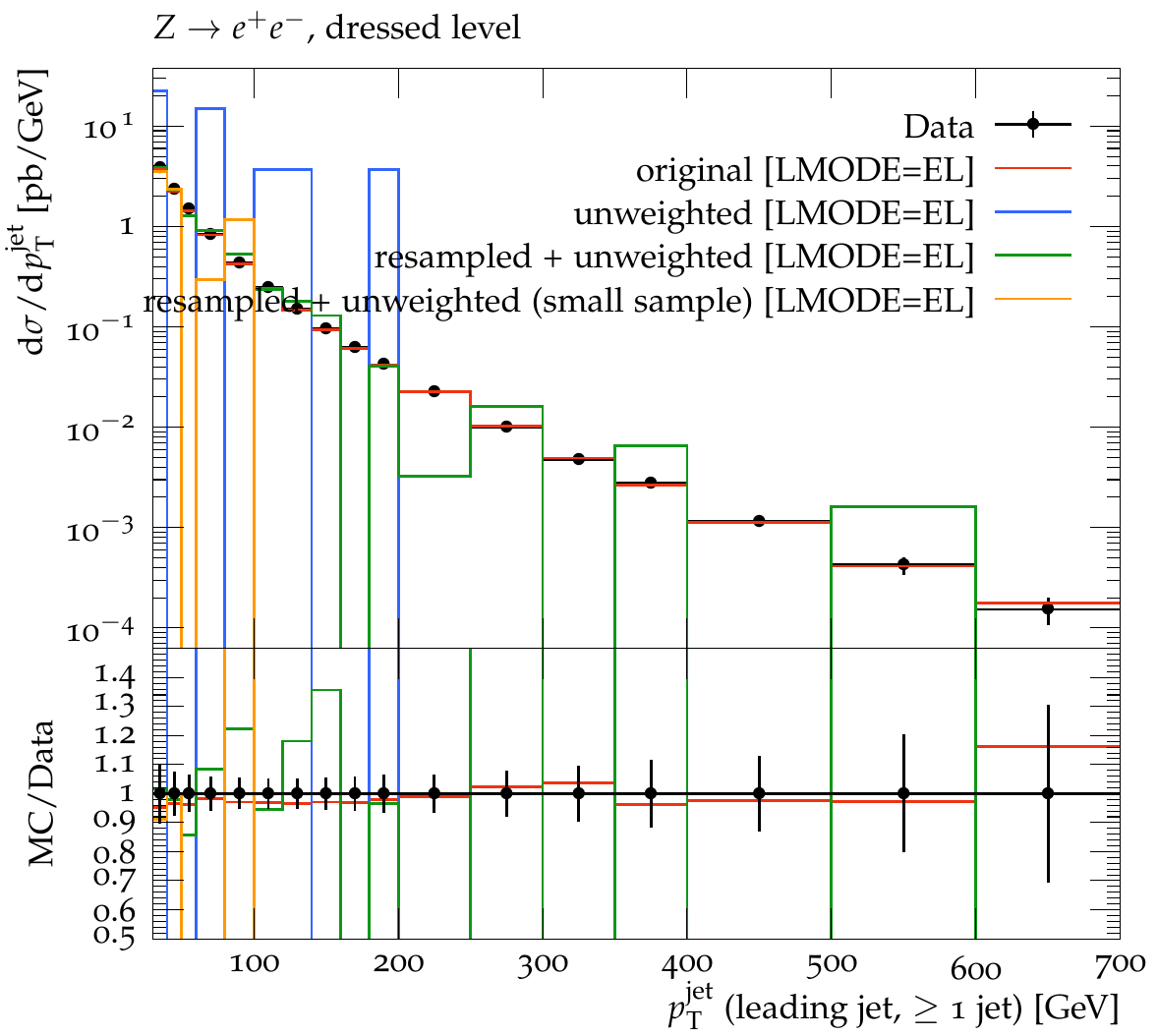}
      \includegraphics[width=0.45\linewidth]{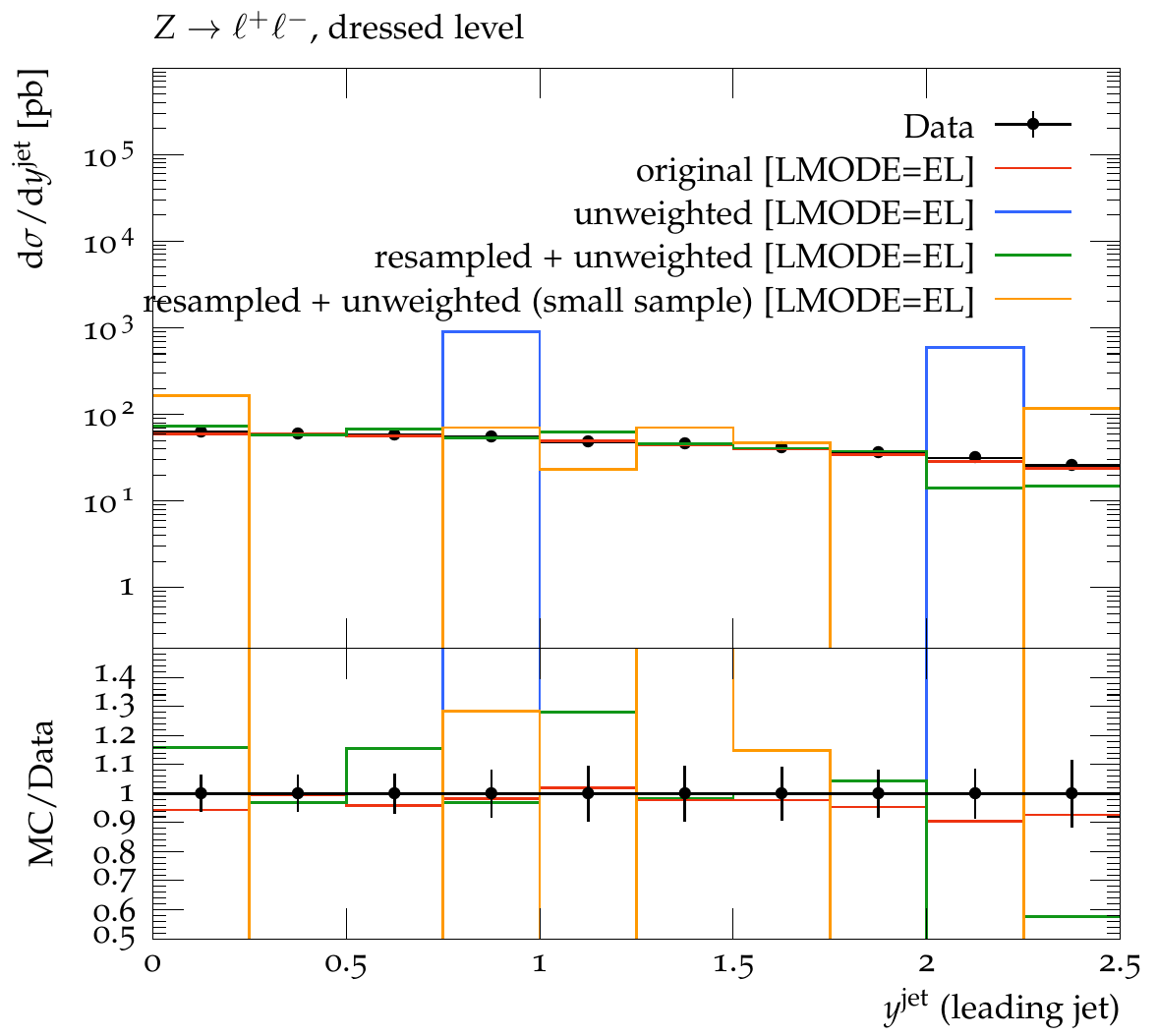}
      \caption{%
        Comparison between unweighted samples before and after
        cell resampling. Lines labeled ``original'' show the reference
        prediction from the original weighted event sample \texttt{Z1}. After
        standard unweighting, the lines with the label ``unweighted'' are
        obtained. Applying cell resampling followed by standard
        unweighting to the sample \texttt{Z1} yields a sample
        represented by the ``resampled + unweighted'' lines. Out of
        this sample, we randomly select a subset matching the size of the
        original ``unweighted'' sample. This leads to the ``resampled +
        unweighted (small sample)'' lines.  Data points are taken
        from~\cite{ATLAS:2017sag}.
      }
  \label{fig:unweight}
\end{figure}

Cell resampling not only reduces the amount of negative weights, but
as a by-product also results in a narrower weight distribution,
enhancing the unweighting efficiency. Indeed, after standard
unweighting we retain only 320 out of the $8.21\times 10^8$ events in
the \texttt{Z1} sample. If we apply resampling beforehand, unweighting
yields 11574 events. The resulting unweighted sample is not only
larger, but also contains a lower fraction of negative-weight
events. We show the gain in statistical power by selecting a subset of
320 randomly chosen events and compare to the unweighted sample based
on the original events. Selected distributions are shown in figure~\ref{fig:unweight}.

\subsection{Runtime Requirements}
\label{sec:timings}

Cell resampling with the improvements presented in
section~\ref{sec:improvements} and a maximum cell size of 10\,GeV
typically takes a few hours of wall-clock time for samples with about
a billion events. As an example, let us consider the resampling for
the \texttt{W5} sample listed in table~\ref{tab:samples}. The combined
size of the original compressed event files is approximately
150\,GB. The resampling program requires about 450\,GB of memory and
the total runtime is about 9 hours on a machine with 24 Intel Xeon
E5-2643 processors. The memory usage could of course be reduced
significantly at the cost of computing time by not keeping all events
in memory, but we have not explored this option in our current
implementation. Reading in the events and converting them to a
space-efficient internal format that only retains the information
needed for resampling takes about 2 hours. This is followed by
approximately 30 minutes spent for the pre-partitioning of the event
sample and less than 3 hours for resampling itself, including the
construction of the search trees,
cf. section~\ref{sec:nearest-neighbour_search}. Since the event
information in the internal format is incomplete, we finally read in
the original events again and write them to disk after updating their
weights. This final step takes roughly 4 hours. While input and output
do not benefit from parallelisation, the pre-partitioning and the
resampling are performed in parallel and the total CPU time spent is
55 hours.

One important optimisation discussed in
section~\ref{sec:nearest-neighbour_search} is trimming the
nearest-neighbour search according to the maximum cell radius. In
fact, when increasing the allowed radius from 10\,GeV to 100\,GeV the
wall clock time needed for resampling rises to several weeks, with a correponding
increase in total CPU time. Extrapolating from smaller
sample sizes, the expected total required CPU time without any of the
new optimisations would be of the order of 1600 years for the much
simpler process of W boson production with two jets considered
in~\cite{Andersen:2021mvw}.


\section{Conclusions}
\label{sec:conclusions}

We have demonstrated that the fraction of negative event weights in
existing large high-multiplicity samples can be reduced by more than
an order of magnitude, whilst preserving predictions for observables
within statistical uncertainties. Concretely, we have employed the cell
resampling method proposed in~\cite{Andersen:2021mvw} with NLO event
samples for Z boson production with up to three jets
and W boson production with five jets produced with \textsc{Sherpa}
and \textsc{BlackHat}.

For the first time, cell resampling has been applied to samples with
up to several billions of events. This was made possible by
algorithmic improvements leading to a speed-up by several orders of
magnitude. Our updated implementation can be retrieved from
\url{https://cres.hepforge.org/}.

The advances in the development of the cell resampling method
presented in this work pave the way for future applications to processes with
high-multiplicities, in particular including parton showered
predictions. It will be necessary to quantify the uncertainty
introduced by the weight smearing. Variations in the maximum cell size
parameter and different prescriptions for weight redistribution within
a cell can serve as handles to assess this uncertainty. Another
promising avenue for further exploration is the analysis of the
information on weight distribution within phase space collected during
cell resampling. Regions with insufficient Monte Carlo statistics
could be identified by their accumulated negative weight, thereby
guiding the event generation. We leave the investigation of these
questions to future work.

\section*{Acknowledgements}

AM thanks Zahari Kassabov for encouragement to reconsider the use of nearest
neighbour search trees. The work of JRA and DM is supported by the STFC under
grant ST/P001246/1.


\appendix

\section{Improved Search Based on Locality-Sensitive Hashing}
\label{sec:LSH_search}

Locality-sensitive hashing (LSH) is a method for approximate
nearest-neighbour search where points are inserted into a number of
hash tables, with hashes that are calculated from the coordinates in
such a way that nearby points end up inside the same hash table
buckets with high probability. To search for a point's nearest
neighbour, one only checks points that share at least a given number
of hash table buckets. An equivalent formulation in the language of
particle physics is to consider a number of one-dimensional
histograms, where the observables are chosen such that similar events
end up in the same histogram bins. To find events that are nearby in
phase space, one only checks those events that share a large number of
histogram bins. A first LSH-based search algortihm for cell resampling
was proposed in~\cite{Andersen:2021mvw}. In the following, we discuss
an improved version, where the histogram observables have a closer
relation to the exact distance measure.

The first step in defining the locality-sensitive observables is the
same as in the exact distance calculation: we cluster the outgoing
particles in each event into infrared-safe physics objects and group
them according to their types. As usual, we add
objects with vanishing momentum to ensure that all events have the
same number of objects for each type. For each object type $t$, we then choose a
random axis $a_t$ in three-dimensional Euclidean space. We choose a final
axis $A$ in a Euclidean space whose dimension is equal to the total number
of infrared-safe physics objects in an event.

For a given event, we then calculate the observable as follows. For
each object type $t$, we project the spatial momentum of each object
onto the previously chosen axis $a_t$ and sort the resulting
coordinates. We concatenate all coordinates obtained in this way into
a single vector. Finally, we obtain the observable by projecting this
vector onto the axis $A$.

We find that the LSH-based search based on the present observables
performs significantly better than the original
version~\cite{Andersen:2021mvw}. However, it still suffers from the
same problem. For constant (or at most logarithmically growing)
numbers of histograms and bin sizes we observe that the typical
distance between an event and the approximate nearest neighbour fails
to decrease with a growing sample size. Hence, we mainly focus on the
exact tree-based search presented in
section~\ref{sec:nearest-neighbour_search}.


\section{Cell Sizes}
\label{sec:cell_sizes}

Larger cell sizes naturally lead to stronger smearing effects. Ideally,
all cells should be small compared to the experimental resolution,
which is limited both by the detector and by statistics.

\begin{figure}[htb]
  \centering
  \includegraphics{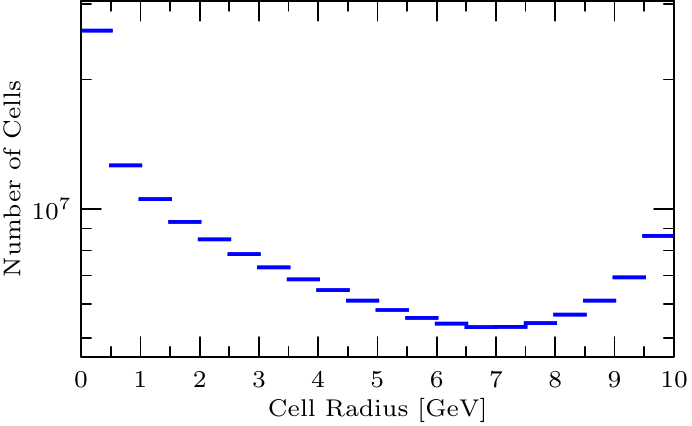}
  \includegraphics{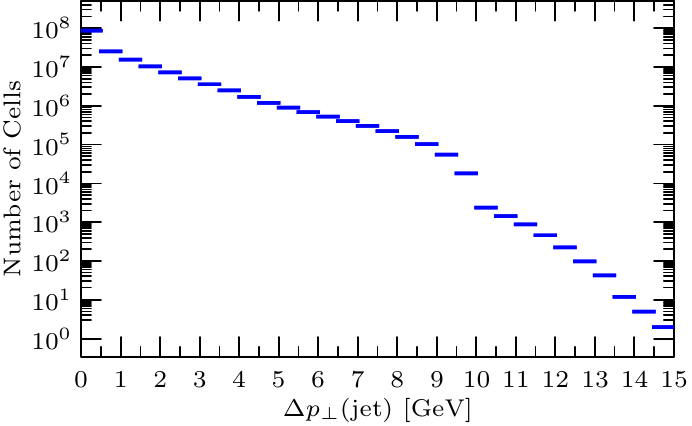}
  \caption{%
    Cell size characteristics for the sample \texttt{Z1}. The left
    pane shows the distribution of cell radii. The right pane displays
    the differences between the transverse momenta of the hardest and
    softest jet within a cell.
  }
  \label{fig:cell_stats}
\end{figure}

In the left pane of figure~\ref{fig:cell_stats} we show the
distribution of cell radii obtained for the \mbox{Z + 1 jet} sample
\texttt{Z1}, cf. table~\ref{tab:samples}. We have omitted cells where
aside from the seed no further event is found within the maximum cell
radius of 10\,GeV. The shape of the distribution is similar to the one
found for W + 2 jets~\cite{Andersen:2021mvw}. The median cell radius
is 3.4\,GeV.

The cell diameter imposes an upper limit on the spread in any single
direction. However, especially in a higher-dimensional phase space, the smearing
range in one-dimensional distributions will be typically much smaller,
as pointed out in our earlier work~\cite{Andersen:2021mvw}. To
illustrate this point, we compute the difference $\Delta
p_\perp(\text{jet})$ between the transverse momenta of the softest jet
and the hardest jet among all events within a cell. The distribution
is shown in the right pane of figure~\ref{fig:cell_stats}. We
observe a steep decline with a median of 0.4\,GeV. There is a notable
drop where $\Delta p_\perp(\text{jet})$ reaches the maximum cell
radius of 10\,GeV. While the theoretical upper limit is given by the
maximum cell diameter of 20\,GeV, we find that the largest transverse
momentum spread in the considered sample is approximately 15\,GeV.


\bibliographystyle{JHEP}
\bibliography{papers}

\providecommand{\href}[2]{#2}\begingroup\raggedright\begin{thebibliography}{10}

\bibitem{HSFPhysicsEventGeneratorWG:2020gxw}
{\scshape HSF Physics Event Generator WG} collaboration, S.~Amoroso et~al.,
  \emph{{Challenges in Monte Carlo Event Generator Software for High-Luminosity
  LHC}}, \href{http://dx.doi.org/10.1007/s41781-021-00055-1}{\emph{Comput.
  Softw. Big Sci.} {\bf 5} (2021) 12},
  [\href{http://arxiv.org/abs/2004.13687}{{\tt 2004.13687}}].

\bibitem{Frixione:1995ms}
S.~Frixione, Z.~Kunszt and A.~Signer, \emph{{Three jet cross-sections to
  next-to-leading order}},
  \href{http://dx.doi.org/10.1016/0550-3213(96)00110-1}{\emph{Nucl. Phys. B}
  {\bf 467} (1996) 399--442}, [\href{http://arxiv.org/abs/hep-ph/9512328}{{\tt
  hep-ph/9512328}}].

\bibitem{Catani:1996vz}
S.~Catani and M.~H. Seymour, \emph{{A General algorithm for calculating jet
  cross-sections in NLO QCD}},
  \href{http://dx.doi.org/10.1016/S0550-3213(96)00589-5}{\emph{Nucl. Phys. B}
  {\bf 485} (1997) 291--419}, [\href{http://arxiv.org/abs/hep-ph/9605323}{{\tt
  hep-ph/9605323}}].

\bibitem{Catani:2002hc}
S.~Catani, S.~Dittmaier, M.~H. Seymour and Z.~Trocsanyi, \emph{{The Dipole
  formalism for next-to-leading order QCD calculations with massive partons}},
  \href{http://dx.doi.org/10.1016/S0550-3213(02)00098-6}{\emph{Nucl. Phys. B}
  {\bf 627} (2002) 189--265}, [\href{http://arxiv.org/abs/hep-ph/0201036}{{\tt
  hep-ph/0201036}}].

\bibitem{Nagy:2003qn}
Z.~Nagy and D.~E. Soper, \emph{{General subtraction method for numerical
  calculation of one loop QCD matrix elements}},
  \href{http://dx.doi.org/10.1088/1126-6708/2003/09/055}{\emph{JHEP} {\bf 09}
  (2003) 055}, [\href{http://arxiv.org/abs/hep-ph/0308127}{{\tt
  hep-ph/0308127}}].

\bibitem{Gehrmann-DeRidder:2005btv}
A.~Gehrmann-De~Ridder, T.~Gehrmann and E.~W.~N. Glover, \emph{{Antenna
  subtraction at NNLO}},
  \href{http://dx.doi.org/10.1088/1126-6708/2005/09/056}{\emph{JHEP} {\bf 09}
  (2005) 056}, [\href{http://arxiv.org/abs/hep-ph/0505111}{{\tt
  hep-ph/0505111}}].

\bibitem{Czakon:2010td}
M.~Czakon, \emph{{A novel subtraction scheme for double-real radiation at
  NNLO}}, \href{http://dx.doi.org/10.1016/j.physletb.2010.08.036}{\emph{Phys.
  Lett. B} {\bf 693} (2010) 259--268},
  [\href{http://arxiv.org/abs/1005.0274}{{\tt 1005.0274}}].

\bibitem{Somogyi:2005xz}
G.~Somogyi, Z.~Trocsanyi and V.~Del~Duca, \emph{{Matching of singly- and
  doubly-unresolved limits of tree-level QCD squared matrix elements}},
  \href{http://dx.doi.org/10.1088/1126-6708/2005/06/024}{\emph{JHEP} {\bf 06}
  (2005) 024}, [\href{http://arxiv.org/abs/hep-ph/0502226}{{\tt
  hep-ph/0502226}}].

\bibitem{Somogyi:2006da}
G.~Somogyi, Z.~Trocsanyi and V.~Del~Duca, \emph{{A Subtraction scheme for
  computing QCD jet cross sections at NNLO: Regularization of doubly-real
  emissions}},
  \href{http://dx.doi.org/10.1088/1126-6708/2007/01/070}{\emph{JHEP} {\bf 01}
  (2007) 070}, [\href{http://arxiv.org/abs/hep-ph/0609042}{{\tt
  hep-ph/0609042}}].

\bibitem{Gaunt:2015pea}
J.~Gaunt, M.~Stahlhofen, F.~J. Tackmann and J.~R. Walsh, \emph{{N-jettiness
  Subtractions for NNLO QCD Calculations}},
  \href{http://dx.doi.org/10.1007/JHEP09(2015)058}{\emph{JHEP} {\bf 09} (2015)
  058}, [\href{http://arxiv.org/abs/1505.04794}{{\tt 1505.04794}}].

\bibitem{Cacciari:2015jma}
M.~Cacciari, F.~A. Dreyer, A.~Karlberg, G.~P. Salam and G.~Zanderighi,
  \emph{{Fully Differential Vector-Boson-Fusion Higgs Production at
  Next-to-Next-to-Leading Order}},
  \href{http://dx.doi.org/10.1103/PhysRevLett.115.082002}{\emph{Phys. Rev.
  Lett.} {\bf 115} (2015) 082002}, [\href{http://arxiv.org/abs/1506.02660}{{\tt
  1506.02660}}].

\bibitem{Bonciani:2015sha}
R.~Bonciani, S.~Catani, M.~Grazzini, H.~Sargsyan and A.~Torre, \emph{{The $q_T$
  subtraction method for top quark production at hadron colliders}},
  \href{http://dx.doi.org/10.1140/epjc/s10052-015-3793-y}{\emph{Eur. Phys. J.
  C} {\bf 75} (2015) 581}, [\href{http://arxiv.org/abs/1508.03585}{{\tt
  1508.03585}}].

\bibitem{Magnea:2018hab}
L.~Magnea, E.~Maina, G.~Pelliccioli, C.~Signorile-Signorile, P.~Torrielli and
  S.~Uccirati, \emph{{Local analytic sector subtraction at NNLO}},
  \href{http://dx.doi.org/10.1007/JHEP12(2018)107}{\emph{JHEP} {\bf 12} (2018)
  107}, [\href{http://arxiv.org/abs/1806.09570}{{\tt 1806.09570}}].

\bibitem{Frixione:2002ik}
S.~Frixione and B.~R. Webber, \emph{{Matching NLO QCD computations and parton
  shower simulations}},
  \href{http://dx.doi.org/10.1088/1126-6708/2002/06/029}{\emph{JHEP} {\bf 06}
  (2002) 029}, [\href{http://arxiv.org/abs/hep-ph/0204244}{{\tt
  hep-ph/0204244}}].

\bibitem{Frixione:2007vw}
S.~Frixione, P.~Nason and C.~Oleari, \emph{{Matching NLO QCD computations with
  Parton Shower simulations: the POWHEG method}},
  \href{http://dx.doi.org/10.1088/1126-6708/2007/11/070}{\emph{JHEP} {\bf 11}
  (2007) 070}, [\href{http://arxiv.org/abs/0709.2092}{{\tt 0709.2092}}].

\bibitem{Hamilton:2012np}
K.~Hamilton, P.~Nason and G.~Zanderighi, \emph{{MINLO: Multi-Scale Improved
  NLO}}, \href{http://dx.doi.org/10.1007/JHEP10(2012)155}{\emph{JHEP} {\bf 10}
  (2012) 155}, [\href{http://arxiv.org/abs/1206.3572}{{\tt 1206.3572}}].

\bibitem{Hoche:2014uhw}
S.~H\"oche, Y.~Li and S.~Prestel, \emph{{Drell-Yan lepton pair production at
  NNLO QCD with parton showers}},
  \href{http://dx.doi.org/10.1103/PhysRevD.91.074015}{\emph{Phys. Rev. D} {\bf
  91} (2015) 074015}, [\href{http://arxiv.org/abs/1405.3607}{{\tt 1405.3607}}].

\bibitem{Jadach:2015mza}
S.~Jadach, W.~P\l{}aczek, S.~Sapeta, A.~Si\'odmok and M.~Skrzypek,
  \emph{{Matching NLO QCD with parton shower in Monte Carlo scheme
  \textemdash{} the KrkNLO method}},
  \href{http://dx.doi.org/10.1007/JHEP10(2015)052}{\emph{JHEP} {\bf 10} (2015)
  052}, [\href{http://arxiv.org/abs/1503.06849}{{\tt 1503.06849}}].

\bibitem{Monni:2019whf}
P.~F. Monni, P.~Nason, E.~Re, M.~Wiesemann and G.~Zanderighi,
  \emph{{MiNNLO$_{PS}$: a new method to match NNLO QCD to parton showers}},
  \href{http://dx.doi.org/10.1007/JHEP05(2020)143}{\emph{JHEP} {\bf 05} (2020)
  143}, [\href{http://arxiv.org/abs/1908.06987}{{\tt 1908.06987}}].

\bibitem{Prestel:2021vww}
S.~Prestel, \emph{{Matching N3LO QCD calculations to parton showers}},
  \href{http://dx.doi.org/10.1007/JHEP11(2021)041}{\emph{JHEP} {\bf 11} (2021)
  041}, [\href{http://arxiv.org/abs/2106.03206}{{\tt 2106.03206}}].

\bibitem{Catani:2001cc}
S.~Catani, F.~Krauss, R.~Kuhn and B.~R. Webber, \emph{{QCD matrix elements +
  parton showers}},
  \href{http://dx.doi.org/10.1088/1126-6708/2001/11/063}{\emph{JHEP} {\bf 11}
  (2001) 063}, [\href{http://arxiv.org/abs/hep-ph/0109231}{{\tt
  hep-ph/0109231}}].

\bibitem{Lonnblad:2001iq}
L.~Lonnblad, \emph{{Correcting the color dipole cascade model with fixed order
  matrix elements}},
  \href{http://dx.doi.org/10.1088/1126-6708/2002/05/046}{\emph{JHEP} {\bf 05}
  (2002) 046}, [\href{http://arxiv.org/abs/hep-ph/0112284}{{\tt
  hep-ph/0112284}}].

\bibitem{Frederix:2012ps}
R.~Frederix and S.~Frixione, \emph{{Merging meets matching in MC@NLO}},
  \href{http://dx.doi.org/10.1007/JHEP12(2012)061}{\emph{JHEP} {\bf 12} (2012)
  061}, [\href{http://arxiv.org/abs/1209.6215}{{\tt 1209.6215}}].

\bibitem{Lonnblad:2012ix}
L.~L\"onnblad and S.~Prestel, \emph{{Merging Multi-leg NLO Matrix Elements with
  Parton Showers}},
  \href{http://dx.doi.org/10.1007/JHEP03(2013)166}{\emph{JHEP} {\bf 03} (2013)
  166}, [\href{http://arxiv.org/abs/1211.7278}{{\tt 1211.7278}}].

\bibitem{Frederix:2020trv}
R.~Frederix, S.~Frixione, S.~Prestel and P.~Torrielli, \emph{{On the reduction
  of negative weights in MC@NLO-type matching procedures}},
  \href{http://arxiv.org/abs/2002.12716}{{\tt 2002.12716}}.

\bibitem{Gao:2020vdv}
C.~Gao, J.~Isaacson and C.~Krause, \emph{{i-flow: High-dimensional Integration
  and Sampling with Normalizing Flows}},
  \href{http://dx.doi.org/10.1088/2632-2153/abab62}{\emph{Mach. Learn. Sci.
  Tech.} {\bf 1} (2020) 045023}, [\href{http://arxiv.org/abs/2001.05486}{{\tt
  2001.05486}}].

\bibitem{Bothmann:2020ywa}
E.~Bothmann, T.~Jan\ss{}en, M.~Knobbe, T.~Schmale and S.~Schumann,
  \emph{{Exploring phase space with Neural Importance Sampling}},
  \href{http://dx.doi.org/10.21468/SciPostPhys.8.4.069}{\emph{SciPost Phys.}
  {\bf 8} (2020) 069}, [\href{http://arxiv.org/abs/2001.05478}{{\tt
  2001.05478}}].

\bibitem{Gao:2020zvv}
C.~Gao, S.~H\"oche, J.~Isaacson, C.~Krause and H.~Schulz, \emph{{Event
  Generation with Normalizing Flows}},
  \href{http://dx.doi.org/10.1103/PhysRevD.101.076002}{\emph{Phys. Rev. D} {\bf
  101} (2020) 076002}, [\href{http://arxiv.org/abs/2001.10028}{{\tt
  2001.10028}}].

\bibitem{Danziger:2021xvr}
K.~Danziger, S.~H\"oche and F.~Siegert, \emph{{Reducing negative weights in
  Monte Carlo event generation with Sherpa}},
  \href{http://arxiv.org/abs/2110.15211}{{\tt 2110.15211}}.

\bibitem{Andersen:2020sjs}
J.~R. Andersen, C.~G\"utschow, A.~Maier and S.~Prestel, \emph{{A Positive
  Resampler for Monte Carlo events with negative weights}},
  \href{http://dx.doi.org/10.1140/epjc/s10052-020-08548-w}{\emph{Eur. Phys. J.
  C} {\bf 80} (2020) 1007}, [\href{http://arxiv.org/abs/2005.09375}{{\tt
  2005.09375}}].

\bibitem{Nachman:2020fff}
B.~Nachman and J.~Thaler, \emph{{Neural resampler for Monte Carlo reweighting
  with preserved uncertainties}},
  \href{http://dx.doi.org/10.1103/PhysRevD.102.076004}{\emph{Phys. Rev. D} {\bf
  102} (2020) 076004}, [\href{http://arxiv.org/abs/2007.11586}{{\tt
  2007.11586}}].

\bibitem{Verheyen:2020bjw}
B.~Stienen and R.~Verheyen, \emph{{Phase Space Sampling and Inference from
  Weighted Events with Autoregressive Flows}},
  \href{http://arxiv.org/abs/2011.13445}{{\tt 2011.13445}}.

\bibitem{Andersen:2021mvw}
J.~R. Andersen and A.~Maier, \emph{{Unbiased elimination of negative weights in
  Monte Carlo samples}},
  \href{http://dx.doi.org/10.1140/epjc/s10052-022-10372-3}{\emph{Eur. Phys. J.
  C} {\bf 82} (2022) 433}, [\href{http://arxiv.org/abs/2109.07851}{{\tt
  2109.07851}}].

\bibitem{Indyk1998}
P.~Indyk and R.~Motwani, \emph{Approximate nearest neighbors: Towards removing
  the curse of dimensionality},  in \emph{Proceedings of the 30th ACM Symposium
  on Theory of Computing}, p.~604–613, 1998.

\bibitem{Leskovec:2020}
J.~Leskovec, A.~Rajaraman and J.~Ullman, \emph{Mining of Massive Datasets}.
\newblock Cambridge University Press, 2020.

\bibitem{UHLMANN1991175}
J.~K. Uhlmann, \emph{Satisfying general proximity / similarity queries with
  metric trees},
  \href{http://dx.doi.org/https://doi.org/10.1016/0020-0190(91)90074-R}{\emph{Information
  Processing Letters} {\bf 40} (1991) 175--179}.

\bibitem{10.5555/313559.313789}
P.~N. Yianilos, \emph{Data structures and algorithms for nearest neighbor
  search in general metric spaces},  in \emph{Proceedings of the Fourth Annual
  ACM-SIAM Symposium on Discrete Algorithms}, SODA '93, (USA), p.~311–321,
  Society for Industrial and Applied Mathematics, 1993.

\bibitem{Jacobi1}
C.~G.~J. Jacobi, \emph{De investigando ordine systematis aequationum
  differentialum vulgarium cujuscunque},  in \emph{C.G.J. Jacobi's gesammelte
  Werke, fünfter Band} (K.~Weierstrass, ed.), pp.~193--216.
\newblock Verlag Georg Reimer, 1890.

\bibitem{Jacobi2}
C.~G.~J. Jacobi, \emph{De aequationum differentialum systemate non normali ad
  formam normalem revocando},  in \emph{C.G.J. Jacobi's gesammelte Werke,
  fünfter Band} (K.~Weierstrass, ed.), pp.~485--513.
\newblock Verlag Georg Reimer, 1890.

\bibitem{Kuhn1}
H.~W. Kuhn, \emph{{The Hungarian Method for the assignment problem}},
  {\emph{Naval Research Logistics Quarterly} {\bf 2} (1955) 83--97}.

\bibitem{Kuhn2}
H.~W. Kuhn, \emph{{Variants of the Hungarian method for assignment problems}},
  {\emph{Naval Research Logistics Quarterly} {\bf 3} (1956) 253--258}.

\bibitem{Munkres}
J.~Munkres, \emph{Algorithms for the assignment and transportation problems},
  {\emph{Journal of the Society for Industrial and Applied Mathematics} {\bf 5}
  (1957) 32--38}.

\bibitem{10.1145/321694.321699}
J.~Edmonds and R.~M. Karp, \emph{Theoretical improvements in algorithmic
  efficiency for network flow problems},
  \href{http://dx.doi.org/10.1145/321694.321699}{\emph{J. ACM} {\bf 19} (apr,
  1972) 248–264}.

\bibitem{https://doi.org/10.1002/net.3230010206}
N.~Tomizawa, \emph{On some techniques useful for solution of transportation
  network problems},
  \href{http://dx.doi.org/https://doi.org/10.1002/net.3230010206}{\emph{Networks}
  {\bf 1} (1971) 173--194},
  [\href{http://arxiv.org/abs/https://onlinelibrary.wiley.com/doi/pdf/10.1002/net.3230010206}{{\tt
  https://onlinelibrary.wiley.com/doi/pdf/10.1002/net.3230010206}}].

\bibitem{pathfinding}
S.~Tardieu, ``pathfinding.'' \url{https://crates.io/crates/pathfinding}, 2022.

\bibitem{Ramshaw2012581}
L.~Ramshaw and R.~E. Tarjan, \emph{A weight-scaling algorithm for min-cost
  imperfect matchings in bipartite graphs},  p.~581 – 590, 2012.
\newblock \href{http://dx.doi.org/10.1109/FOCS.2012.9}{DOI}.

\bibitem{Bertsekas1988}
D.~P. Bertsekas, \emph{Auction algorithms for network flow problems: A tutorial
  introduction}, {\emph{Computational optimization and applications} {\bf 1}
  (1992) 7–66}.

\bibitem{Goldberg1995}
A.~Goldberg and R.~Kennedy, \emph{An efficient cost scaling algorithm for the
  assignment problem},
  \href{http://dx.doi.org/10.1007/BF01585996}{\emph{Mathematical Programming}
  {\bf 71} (03, 1995) 153--177}.

\bibitem{Alfaro2022}
C.~Alfaro, S.~Perez~Perez, C.~Valencia and M.~Vargas~Magaña, \emph{The
  assignment problem revisited},
  \href{http://dx.doi.org/10.1007/s11590-021-01791-4}{\emph{Optimization
  Letters} {\bf 16} (06, 2022) }.

\bibitem{Brun:1997pa}
R.~Brun and F.~Rademakers, \emph{{ROOT: An object oriented data analysis
  framework}},
  \href{http://dx.doi.org/10.1016/S0168-9002(97)00048-X}{\emph{Nucl. Instrum.
  Meth. A} {\bf 389} (1997) 81--86}.

\bibitem{Bern:2013zja}
Z.~Bern, L.~J. Dixon, F.~Febres~Cordero, S.~H\"oche, H.~Ita, D.~A. Kosower
  et~al., \emph{{Ntuples for NLO Events at Hadron Colliders}},
  \href{http://dx.doi.org/10.1016/j.cpc.2014.01.011}{\emph{Comput. Phys.
  Commun.} {\bf 185} (2014) 1443--1460},
  [\href{http://arxiv.org/abs/1310.7439}{{\tt 1310.7439}}].

\bibitem{Anger:2017nkq}
F.~R. Anger, F.~Febres~Cordero, S.~H\"oche and D.~Ma\^\i{}tre, \emph{{Weak
  vector boson production with many jets at the LHC $\sqrt{s}= 13$ TeV}},
  \href{http://dx.doi.org/10.1103/PhysRevD.97.096010}{\emph{Phys. Rev. D} {\bf
  97} (2018) 096010}, [\href{http://arxiv.org/abs/1712.08621}{{\tt
  1712.08621}}].

\bibitem{Berger:2008sj}
C.~F. Berger, Z.~Bern, L.~J. Dixon, F.~Febres~Cordero, D.~Forde, H.~Ita et~al.,
  \emph{{An Automated Implementation of On-Shell Methods for One-Loop
  Amplitudes}}, \href{http://dx.doi.org/10.1103/PhysRevD.78.036003}{\emph{Phys.
  Rev. D} {\bf 78} (2008) 036003}, [\href{http://arxiv.org/abs/0803.4180}{{\tt
  0803.4180}}].

\bibitem{Gleisberg:2008ta}
T.~Gleisberg, S.~Hoeche, F.~Krauss, M.~Schonherr, S.~Schumann, F.~Siegert
  et~al., \emph{{Event generation with SHERPA 1.1}},
  \href{http://dx.doi.org/10.1088/1126-6708/2009/02/007}{\emph{JHEP} {\bf 02}
  (2009) 007}, [\href{http://arxiv.org/abs/0811.4622}{{\tt 0811.4622}}].

\bibitem{Cacciari:2008gp}
M.~Cacciari, G.~P. Salam and G.~Soyez, \emph{{The anti-$k_t$ jet clustering
  algorithm}},
  \href{http://dx.doi.org/10.1088/1126-6708/2008/04/063}{\emph{JHEP} {\bf 04}
  (2008) 063}, [\href{http://arxiv.org/abs/0802.1189}{{\tt 0802.1189}}].

\bibitem{Bern:2013gka}
Z.~Bern, L.~J. Dixon, F.~Febres~Cordero, S.~H\"oche, H.~Ita, D.~A. Kosower
  et~al., \emph{{Next-to-Leading Order $W + 5$-Jet Production at the LHC}},
  \href{http://dx.doi.org/10.1103/PhysRevD.88.014025}{\emph{Phys. Rev. D} {\bf
  88} (2013) 014025}, [\href{http://arxiv.org/abs/1304.1253}{{\tt 1304.1253}}].

\bibitem{Bierlich:2019rhm}
C.~Bierlich et~al., \emph{{Robust Independent Validation of Experiment and
  Theory: Rivet version 3}},
  \href{http://dx.doi.org/10.21468/SciPostPhys.8.2.026}{\emph{SciPost Phys.}
  {\bf 8} (2020) 026}, [\href{http://arxiv.org/abs/1912.05451}{{\tt
  1912.05451}}].

\bibitem{ATLAS:2017sag}
{\scshape ATLAS} collaboration, M.~Aaboud et~al., \emph{{Measurements of the
  production cross section of a $Z$ boson in association with jets in pp
  collisions at $\sqrt{s} = 13$ TeV with the ATLAS detector}},
  \href{http://dx.doi.org/10.1140/epjc/s10052-017-4900-z}{\emph{Eur. Phys. J.
  C} {\bf 77} (2017) 361}, [\href{http://arxiv.org/abs/1702.05725}{{\tt
  1702.05725}}].

\end{thebibliography}\endgroup

\end{document}